\documentclass[11pt]{article}

\usepackage[utf8]{inputenc}
\usepackage[top=1in,bottom=1in,left=0.75in,right=1in]{geometry}
\usepackage{amssymb,amsmath}
\usepackage{feynmp}
\usepackage[force]{feynmp-auto}
\usepackage{esvect}
\usepackage{slashed}
\usepackage{graphicx}
\usepackage{dsfont}
\usepackage{cancel}
\usepackage{ulem}

\newcommand{\be}{\begin{eqnarray}}
\newcommand{\ee}{\end{eqnarray}}
\newcommand{\id}{{\mathbb I}}

\begin{document}
\begin{fmffile}{diagram}

\title{One-loop same helicity YM amplitudes from BG currents}
\author{Pratik Chattopadhyay and Kirill Krasnov\\ {}\\
{\it School of Mathematical Sciences, University of Nottingham, NG7 2RD, UK}}

\date{October 2021}
\maketitle
\begin{abstract} We propose and prove a new formula for the one-loop all same helicity Yang-Mills amplitudes. These amplitudes are seen to arise as a sum of products of two tree-level Berends-Giele currents connected by an effective propagator. To make sense of the propagators one needs to introduce the so-called region, or dual momenta. The formula is proven by observing that it readily implies the correct collinear limit properties. The only non-trivial part of the proof is establishing that our formula for the amplitude is invariant under shifts of the region momenta.  
\end{abstract}

\section{Introduction}

The purpose of this paper is to propose a new formula for the same helicity one-loop Yang-Mills amplitudes. A general formula for such amplitudes is known from \cite{Bern:1993sx}, \cite{Bern:1993qk}, where it was conjectured and then proven by checking that it satisfies the required collinear properties. An alternative derivation follows from the results of \cite{Mahlon:1993fe}.

The structure of the formula we propose is rather different from that in \cite{Bern:1993qk}. We show that the same helicity one-loop YM amplitudes can be built from the same helicity Berends-Giele (tree-level) off-shell currents connected by certain effective propagators. Importantly, the propagators can only be made sense of if one introduces a parametrisation in terms of region momenta. 

The new expression for the one-loop amplitudes is as follows. Consider a colour ordered diagram with n external lines. Because the diagram is colour ordered we can adopt the use of region momenta $p_i$ so that the null momenta on the external legs are given by the difference of the region momenta 
 \be 
 k_i=p_{i}-p_{i-1}, ~~~p_n\equiv p_0 .
 \ee 
 This ensures momentum conservation, but introduces indeterminacy in that all region momenta can be shifted by an arbitrary amount without changing the external momenta.
 
The amplitude can then be written as a sum over partitions of the set of external states into two groups. Each group of states has a Berends-Giele (BG) \cite{Berends:1987me} current associated with it, and two such currents are glued by a certain effective propagator. Berends-Giele currents are vectorial objects, but their vector structure is simple and universal, see (\ref{BG-vector}) below. Let $J(1,\ldots, n)$ be the most interesting scalar part of the current given in terms of the momentum spinors as well as the reference spinor, see (\ref{YM-BG}). The one-loop amplitude can then be written pictorially as follows

 \be \label{A-YM-general}
{\cal A}= \sum_{part} \begin{gathered}
~~~\begin{fmfgraph*}(120,70)
     \fmfleft{i,i1,i2,i3,i4}
     \fmfright{o1,o4,o5,o6,o7}
     \fmf{vanilla}{i,v}
     \fmf{vanilla}{i4,v}
     \fmfblob{.15w}{v}
     \fmf{vanilla,label=$p_j$,label.distance=1cm,label.side=left}{v,v1}
     \fmfdot{v1}
     \fmf{vanilla,label=$p_{i-1}$,label.distance=1cm,label.side=right}{v1,o}
     \fmfblob{.15w}{o}
     \fmf{vanilla}{o,o1}
     \fmf{vanilla}{o,o7}
     \fmflabel{$i$}{i}
      \fmflabel{$.$}{i1}
     \fmflabel{$.$}{i2}
      \fmflabel{$.$}{i3}
      \fmflabel{$j$}{i4}
     \fmflabel{$i-1$}{o1}
      \fmflabel{$.$}{o4}
      \fmflabel{$.$}{o5}
      \fmflabel{$.$}{o6}
     \fmflabel{$j+1$}{o7}    
     \end{fmfgraph*}
     \end{gathered}
     \quad\quad = \sum_{part} J(i,..,j)J(j+1,..,i-1)\langle q|p_j\circ p_{i-1} |q\rangle^2.
\ee 
The large blobs in the pictorial representation of the amplitude stand for the BG currents, with the off-shell legs pointing towards the internal line of this diagram. The other quantities are as follows. The object $q$ is the reference (auxiliary) spinor. The sum is taken over partitions of a cyclically ordered set $1,\ldots, n$ with the convention $n+1=1$ into two subgroups $i,\ldots, j$ and $j+1,\ldots, i-1$. The bra-ket notation is standard and represents a particular 2-spinor contraction. The notation $p\circ q$ represents a contraction of two rank two spinors $(p\circ q)_{M}{}^{N} := p_M{}^{M'} q_{M'}{}^{N}$, so that the result can be inserted between two copies of the unprimed reference spinor $q_M$. The momenta $p_j, p_{i-1}$ are the region momenta for the two regions that are separated by the internal line that partitions the two groups of external states. It is thus clear that the object $\langle q|p_j\circ p_{i-1} |q\rangle^2$ plays the role of the effective propagator connecting the BG currents, and that this propagator can only be made sense of if region momenta are introduced. There is also a version of our formula that is written in terms of the vector-valued BG currents that are connected by an effective gluon propagator. This will be spelled out in the main text. 

The formula (\ref{A-YM-general}) is proven by establishing that it has the correct collinear properties of a one-loop same helicity amplitude. This follows quite easily from the construction of the amplitude if one assumes that the right-hand side of (\ref{A-YM-general}) is invariant under shifts of all region momenta by the same amount. So, the difficulty of the proof of the formula (\ref{A-YM-general}) shifts to the proof that it is independent of a choice of the region momenta. These arguments are given in Section \ref{sec:checks}. We also explicitly check that the new formula gives the correct known expressions in the case of four and five particles. 

There are several previous works that are strongly related to the context of this work. First, Zvi Bern (unpublished) observed that in the computation of the same helicity one-loop amplitude in Yang-Mills theory, the sum of box, triangle and bubble integrands, with a particular choice of the loop momenta, vanishes. We reproduce this argument in the Appendix, in our notations, for completeness. This has later been explored by \cite{Chakrabarti:2005ny} in the context of world-sheet formulation of YM, and by \cite{Brandhuber:2007vm} in the related light-cone formalism. There is also a very closely related discussion, again in the light-cone, in the context of chiral higher spin theories in \cite{Skvortsov:2020gpn}. The upshot of these works is that the same helicity one-loop amplitudes in YM can be obtained by performing tree-level calculations, but with insertions of certain helicity violating "bubble" counterterms. Our new formula (\ref{A-YM-general}) is the same observation phrased in the covariant formalism. In particular, we will see that the effective propagator in (\ref{A-YM-general}) arises as the result of a "bubble" computation. 

The organisation of the rest of this paper is as follows. We start by reviewing in Section \ref{sec:SDYM} the covariant Feynman rules of the self-dual Yang-Mills theory. This theory is a truncation of the full YM, and has simpler Feynman rules. At the same time the amplitudes that are of interest to us coincide in the full and SDYM, and so it is sufficient to work with an easier truncation. We also review in this section the by now standard results on the YM BG currents, and also review the previously known formulas for the same helicity one-loop amplitude. Section \ref{sec:bubble} computes the bubble diagram and thus extracts the form of the effective propagator to be used in the construction of the one-loop amplitude. We do checks of the new formula in Section \ref{sec:checks}. We end with a discussion. Appendix spells out the details of the computation that shows that the sum of integrands of the box, triangle and bubble diagrams, with a specific consistent choice of the loop momentum, vanishes. This shows that the 4-point amplitude can be re-written in terms of the bubble insertions. Our formula (\ref{A-YM-general}) is the generalisation of this statement to an arbitrary number of external gluons.

\section{Review of SDYM theory}
\label{sec:SDYM}

Self-dual Yang-Mills is a truncation of the full YM theory that keeps only a subset of the Feynman diagrams of the latter. For some processes, for example those involving mostly gluons of same helicity, only the Feynman diagrams already present in SDYM can be seen to contribute in the full YM calculations. This means that some of the processes can be computed in the simpler SDYM theory. In particular, the same helicity one-loop amplitudes of the full YM are correctly captured by the SD theory. Our presentation in this section follows \cite{Krasnov:2016emc}, where we also direct the reader for more details.

\subsection{Self-dual Yang-Mills} 

The most useful covariant formulation of SDYM is due to Clamers and Siegel \cite{Chalmers:1996rq}. We will state the action in 2-component spinor notations, which are used throughout this paper. Our 2-component spinor conventions are spelled out in the Appendix of \cite{Krasnov:2016emc}. The action reads
\be
S[\phi, A] = \int \Phi^{a\, AB} (\partial_{A}{}^{A'} A^a_{A'B} + f^{abc} A^b_{A}{}^{A'} A^c_{A'B}).
\ee
The field $\Phi^{a\, AB}$ is a symmetric rank two spinor $\Phi^{a\, AB}= \Phi^{a\, (AB)}$ with values in the Lie algebra of some (compact) group $G$, and $a,b,c$ are the Lie algebra indices. The capital Latin letters (unprimed and primed) at the 2-component Lorentz spinor indices. The field $\Phi^{a\, AB}$ is an auxiliary field the variation of the action with respect to which imposes the equation that the self-dual part of the field strength of the connection $A^a_{AA'}$ vanishes. 

The action can be expanded around any self-dual connection configuration. The simplest case is to expand around the zero connection. The first term is then the kinetic term, whose inverse, after an appropriate gauge-fixing, gives the propagator of the theory. The second term is a simple cubic non-derivative interaction. The full YM is obtained from this theory by appending to the action a term quadratic in $\Phi^{a\, AB}$, see e.g. \cite{Krasnov:2016emc}. 

The gauge-fixing of the above action is most usefully done by prolonging the auxiliary field $\Phi^{a\, AB}$ to contain also the $AB$ anti-symmetric part. The variation with respect to this anti-symmetric part imposes the sharp Landau-type gauge $\partial^{AA'} A^a_{AA'}=0$. The gauge-fixed kinetic operator is a version of the (chiral) Dirac operator, is non-degenerate, and can be inverted. The propagator of the theory thus connects the $\Phi$ and $A$ legs of the vertices, and is, schematically, in momentum space $1/k$. The operator in the denominator is the Dirac operator (in the momentum space). Using the fact that Dirac operator squares to the Laplacian, the propagator can be rewritten as $k/k^2$, where the Dirac operator is now in the numerator, and the denominator contains the usual factor of $k^2$. For details on the gauge-fixing procedure, see \cite{Krasnov:2016emc}. 

The variation of the (linearised) action with respect to the auxiliary field gives the field equation $\partial_{A}{}^{A'} A^a_{A'B}=0$. The solutions describe one of the two gluon helicity, in our conventions negative. The corresponding polarisation vector is
\be
\epsilon_-^{MM'}(k)= \frac{q^M k^{M'}}{ \langle kq\rangle}.
\ee
Here $\langle \lambda\nu\rangle = \lambda^M \nu_M$ is the spinor contraction of two 2-component spinors $\lambda^M, \nu^M$. The object $q^M$ is an auxiliary spinor changing which amounts to gauge-transformations. The object $k^{M'}$ is the momentum spinor arising as the square root of a null momentum $k^{MM'} = k^M k^{M'}$. The other gluon helicity is described by the field $\Phi$, but we will never consider processes with this other helicity on the external lines in this paper, and so we do not need to define it. 

\subsection{BG current for all negative helicity states}

The sum of all (colour-ordered) tree-level Feynman diagrams with the negative helicity states inserted on all legs except one, the later being kept off-shell, can be computed explicitly using the Berends-Giele recursion \cite{Berends:1987me}. Using the Feynman rules reviewed in the previous subsection, the computation is very simple and is spelled out in e.g. \cite{Krasnov:2016emc}. The result is as follows. 

Let us start by writing the single particle negative helicity state in the form of a BG current. We have
\be
\epsilon_-^{MM'}(k) = J(k) q^N k_{N}{}^{M'} q^M \equiv J(k) \langle q| \langle q| k| ,
\ee 
where $k_{MM'} = k_M k_{M'}$ is the (null) momentum of the particle and
\be
J(k) = \frac{1}{\langle qk\rangle \langle kq\rangle}.
\ee
The BG current is then defined to be the sum of all (colour-ordered) Feynman diagrams, with negative helicity states of momenta $k_1, \ldots, k_n$ inserted (from the top), and the single remaining off-shell leg (at the bottom). The convention is that the propagator is included in the off-shell leg. It can then be shown by a recursion that all currents have the same form
\be\label{BG-vector}
J^{MM'}(k_1,\ldots, k_n) = J(k_1,\ldots, k_n) q^N (\sum_{i=1}^n k_{i\, N}{}^{M'}) q^M \equiv J(k) \langle q| \langle q| \sum_i k_i | ,
\ee
where
\be\label{YM-BG}
J(k_1,\ldots, k_n) = \frac{1}{\langle q1\rangle \langle 12\rangle \langle 23\rangle \ldots \langle (n-1) n\rangle \langle nq\rangle},
\ee
and we use the convention in which $k_1\equiv 1$ etc. 

These BG currents could be converted to tree-level scattering amplitudes, but all such amplitudes are easily seen to vanish. This is because they need to be multiplied by the factor of $k^2$ for the off-shell leg, which then must be taken on-shell. Thus, all BG currents get multiplied by the factor of zero in the process of converting them to amplitudes. Thus, all tree-level amplitudes in SDYM vanish. 

\subsection{Literature representation for one-loop amplitudes} 

The only non-vanishing amplitudes in SDYM are the one-loop ones, with all external lines being states of negative helicity. The all same helicity one loop amplitudes $A_n(1^+,2^+,...n^+)$ in YM are finite rational functions of the momenta involved and has cyclical symmetry in the arguments. These amplitudes are singular in the region where two adjacent momenta become collinear or a momentum become soft. These amplitudes were explicitly computed at 4 and 5 points in \cite{Bern:1993mq}, \cite{Bern:1993sx}, and then an expression conjectured for them at arbitrary multiplicity. This was then proved in \cite{Bern:1993qk} by showing that the proposed formula has the correct collinear properties of a one-loop amplitude, and also by \cite{Mahlon:1993fe} by an explicit computation in a simpler theory of massless QED. 

The amplitudes at 4 and 5 points are given by 
\be
\label{A4}
\mathcal{A}_4(1^+,2^+,3^+,4^+)=\frac{s_{12}s_{23}}{\langle 12\rangle\langle 23\rangle\langle34\rangle\langle 41\rangle}, \\
\mathcal{A}_5(1^+,2^+,3^+,4^+,5^+)=\frac{s_{12}s_{23}+s_{23}s_{34}+s_{34}s_{45}+s_{45}s_{51}+s_{51}s_{12}+\epsilon(1,2,3,4)}{\langle 12\rangle\langle 23\rangle\langle34\rangle\langle 45\rangle\langle 51\rangle},
\ee 
where $\epsilon(i,j,m,n)=[ij]\langle jm\rangle[mn]\langle ni\rangle-\langle ij\rangle[jm]\langle mn\rangle[ni]$. In this paper we are not interested in the overall multiplicative factors in the amplitudes, and these are omitted. 

There are many other equivalent ways to write the 5-point amplitude. For example, the paper  \cite{Bern:1993sx} contains the following more compact expression
\be\label{A5-altern}
\mathcal{A}_5(1^+,2^+,3^+,4^+,5^+)&={s_{12} s_{23} + s_{45} s_{51} - \langle 23\rangle [34] \langle 45\rangle [52]}.
\ee
The notation that is used everywhere is $s_{ij} = \langle ij\rangle [ij]$. 

The formula that is valid at an arbitrary multiplicity is \cite{Mahlon:1993fe}, \cite{Bern:1993qk}
\be \label{YM-known-expression}
\mathcal{A}_n(1^+,2^+,....n^+)=\sum_{1\leq i_1<i_2<i_3<i_4\leq n}\frac{\langle i_1i_2\rangle[i_2i_3]\langle i_3i_4\rangle[i_4i_1]}{\langle 12\rangle\langle 23\rangle....\langle n1\rangle}.
\ee

\section{Bubble computation and effective propagator}
\label{sec:bubble}

The purpose of this section is to spell out the computation of the one-loop "bubble" correction to the self-energy diagram. We will see that while this diagram can be argued to vanish, it is in fact undefined. If the loop momentum is shifted, the difference between the unshifted and shifted value of this diagram is non-zero an can be computed. Thus, this diagram must in general be assigned a non-zero value. In our formula for the one-loop amplitudes it is this diagram, with a specific prescription for the value of the shifted momentum, that is inserted between the two BG currents. 

\subsection{Shift computation}

~~~~~~~~~~~~~~~~~~~~~~~~~~~~~~~~~
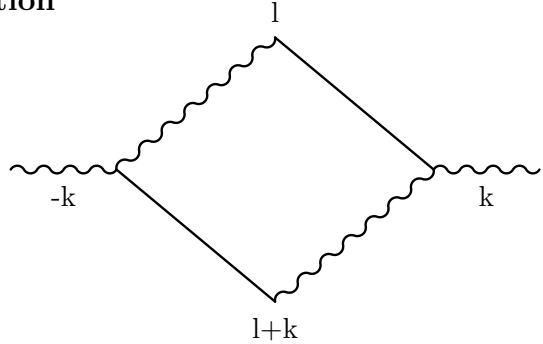
\begin{fmfgraph*}(200,100)
     \fmfleft{i}
     \fmfright{o}
     \fmfbottom{v}
     \fmftop{u}
     \fmf{photon,tension=3,label=-k}{i,v1}
     \fmf{photon}{v1,u}
     \fmf{vanilla}{u,v2}
     \fmf{photon}{v2,v}
     \fmf{vanilla}{v,v1}
     \fmf{photon,tension=3,label=k}{v2,o}
     \fmflabel{l}{u}
     \fmflabel{l+k}{v}
\end{fmfgraph*}
\\~\\
\\~\\
Consider the self-energy diagram as pictured above, where the external lines are projected to two negative helicity states and the convention being all external momenta incoming. Using the Feynman rules reviewed in the previous section, and the bra-ket spinor notation, the amplitude can be written as 
\begin{equation}
\label{se}
i\Pi^{--}=\int \frac{d^4l}{(2\pi)^4}\frac{\langle q|l|k]\langle q|l+k|k]}{l^2(l+k)^2\langle qk\rangle^2}.
\end{equation}
In the form it is written this integral can be argued to vanish. Indeed, 
there is no linear in $l$ part of the integrand as it is proportional to $\langle q|k|k]=\langle qk\rangle[k k]=0$. The only non-vanishing contribution thus comes from 
\be 
\int \frac{d^4l}{(2\pi)^4}\frac{l_{\mu}l_{\nu}}{l^2(l+p)^2}
\ee
Any Lorentz invariant regularisation of this will yield $x_{\mu}x_{\nu}$ to be proportional to $k^2 \eta_{\mu\nu}$ which is zero because $k$ is null or to $k_{\mu}k_{\nu}$, which gives the numerator factor $\langle q|k|k]=0$ by using $[kk]=0$. Thus, the self-energy diagram (\ref{se}) can be argued to be zero. 

However, (\ref{se}) is a quadratically divergent integral, and so one must be careful in reaching the conclusion that this object is zero. Let us consider shifting the loop momentum as in $l= x+ \tilde{s}$, where $\tilde{s}$ is some momentum and $x$ is the new integration variable. The argument above depends on the specific form of the integrand and is no longer applicable to the shifted integrand. In fact, below we shall compute the effect of the shift by $\tilde{s}$ and see that the shift is non-vanishing. What this means is that the self-energy diagram projected onto two negative helicity states cannot in general be assumed to vanish. Instead, it is given by a finite quantity, depending on the shift parameter. 

Let us compute the shift dependence of the self-energy diagram. We will use the region momenta so that $k=s-\tilde{s}$, and 
\be 
\begin{split} 
\label{rel}
l&=x+\tilde{s}\\
l+k&=x + s.
\end{split}
\ee 
We then have
\be 
\label{dse}
i\Pi^{--}=\int \frac{d^4x}{(2\pi)^4}\frac{\langle q|x+\tilde{s}|k]\langle q|x+s|k]}{(x+\tilde{s})^2(x+s)^2\langle qk\rangle^2}
\ee
We have already see that this integral vanishes after the shift $x\rightarrow x-\tilde{s}$. We then compute the result of the shift. This is done using the standard techniques, which are reviewed in the Appendix of \cite{Chattopadhyay:2020oxe}. The linear part of the shift is given by 
\be 
-i\lim_{x\to\infty}\int\frac{d\Omega}{(2\pi)^4}\tilde{s}_{\mu}x^{\mu}\frac{\langle q|x+\tilde{s}|k]\langle q|x+s|k]}{x^2}\Bigg(1-\frac{2x.(\tilde{s}+s)}{x^2}\Bigg)
\ee 
The non-zero contribution can only come from the quadratic and quartic in $x$ terms. The quadratic term is 
\be 
\tilde{s}_{\mu}x^{\mu}\Big(\langle q|x|k]\langle q|s|k]+\langle q|x|k]\langle q|\tilde{s}|k]\Big)
\ee 
Integrating over the directions of $x^\mu$ produces
\be 
-\frac{i}{32\pi^2}\Big(\langle q|\tilde{s}|k]\langle q|s|k]+\langle q|\tilde{s}|k]\langle q|\tilde{s}|k]\Big)
\ee 
The quartic in $x$ part is given by 
\be 
2i\lim_{x\to\infty}\int\frac{d\Omega}{(2\pi)^4}\tilde{s}_{\mu}x^{\mu}(\tilde{s}+s)_{\nu}x^{\nu}\frac{\langle q|x|k]\langle q|x|k]}{x^4}
\ee 
The integral is computed using 
\be
\label{quart-YM}
\int\frac{d\Omega}{(2\pi)^4}\frac{x_{\mu}x_{\nu}x_{\rho}x_{\sigma}}{x^4}=\frac{1}{32.6\pi^2}(\eta_{\mu\nu}\eta_{\rho\sigma}+\eta_{\mu\rho}\eta_{\nu\sigma}+\eta_{\mu\sigma}\eta_{\rho\nu})
\ee 
This results in the following two contributions
\be
\begin{split} 
\frac{i}{32.3\pi^2}\Big(\langle q|\tilde{s}|k]\langle q|\tilde{s}+s|k]+\langle q|\tilde{s}+s|k]\langle q|\tilde{s}|k]\Big)
=\frac{i}{16.3\pi^2}\Big(\langle q|\tilde{s}|k]\langle q|\tilde{s}|k]+\langle q|\tilde{s}|k]\langle q|s|k]\Big)
\end{split}
\ee 
For the quadratic part of the shift, the integral is given by 
\be 
\label{quadratic}
\frac{i}{2}\lim_{x\to\infty}\int\frac{d\Omega}{(2\pi)^4}\tilde{s}_{\mu}\tilde{s}_{\nu}x^{\mu}x^2\frac{\partial}{\partial x_{\nu}}\frac{\langle q|x+\tilde{s}|k]\langle q|x+s|k]}{(x+\tilde{s})^2(x+s)^2}
\ee 
When the derivative hits the denominator, it produces a factor proportional to 
\be 
\frac{i}{2}(-4)\lim_{x\to\infty}\int\frac{d\Omega}{(2\pi)^4}\tilde{s}_{\mu}\tilde{s}_{\nu}x^{\mu}x^{\nu}\frac{\langle q|x+\tilde{s}|k]\langle q|x+s|k]}{x^4}
\ee 
The quartic in $x$ part of the numerator is the only which contributes. Using (\ref{quart-YM}) we find one of the contractions vanish and the other two contractions are equal, giving
\be 
-\frac{2i}{32.3\pi^2}\langle q|\tilde{s}|k]\langle q|\tilde{s}|k]
\ee 
When the derivative in (\ref{quadratic}) hits the numerator, in the large $x$ limit, we get 
\be 
\frac{i}{2}\lim_{x\to\infty}\int\frac{d\Omega}{(2\pi)^4}\tilde{s}_{\mu}x^{\mu}\frac{\langle q|x+\tilde{s}|k]\langle q|\tilde{s}|k]+\langle q|x+s|k]\langle q|\tilde{s}|k]}{x^2}
\ee 
Using the relevant contraction, this gives 
\be 
\frac{i}{32\pi^2}\langle q|\tilde{s}|k]^2
\ee
Adding all the contributions, we have for this amplitude 
\be 
\label{shiftr-ym}
\Pi^{--}= -\frac{i}{32.3\pi^2}\frac{\langle q|\tilde{s}|k]\langle q|s|k]}{\langle qk\rangle^2}.
\ee
Using the fact that $k=s-\tilde{s}$ we could write this result in terms of only $s$ or $\tilde{s}$, but the form we chose will be most convenient below. 

\subsection{Bubble as an effective propagator}

As we have just seen, a direct computation of the shift gives the result (\ref{shiftr-ym}) for the self-energy diagram. This result can be given the following interpretation. We represent it as the effective propagator
\be
\label{eff-prop}
\begin{gathered}
\begin{fmfgraph*}(140,70)
     \fmfleft{i}
     \fmfright{o}
     \fmf{vanilla}{i,v1}
     \fmfdot{v1}
     \fmf{vanilla}{o,v1}
     \fmflabel{$MM'$}{i}
     \fmflabel{$NN'$}{o}
     \end{fmfgraph*}
     \end{gathered}~~~~~~~~~~:=~~~~~s^{~M}_{N'} \tilde{s}_{N}^{~M'},
\ee 
into which the polarisation state 
\be
\epsilon^-_{MM'}(k)=\frac{q_M k_{M'}}{\langle qk\rangle}
\ee
is inserted on both sides. This results in (\ref{shiftr-ym}).

\section{New formula}
\label{sec:checks}

It is now easy to see that the the sum of two copies of the Berends-Giele currents (\ref{BG-vector}) connected by the effective propagator (\ref{eff-prop}) where $s,\tilde{s}$ stand for the region momenta separated by the propagator line, is given by the right-hand side of the formula (\ref{A-YM-general}). We then conjecture that this gives the correct formula for the one-loop same helicity YM amplitudes. A proof of this conjecture, as well as explicit computations in the case of 4 and 5 points are collected in this section. 

\subsection{Collinear limit}

\subsubsection{Review of the collinear limit properties}

In \cite{Bern:1993qk}, the form (\ref{YM-known-expression}) of the one-loop amplitude has been fixed using the requirement that it has the correct collinear limit behaviour. The general form of the collinear limit is, see \cite{Bern:1993qk}
\be 
\label{coloop}
\mathcal{A}^{\textrm{loop}}_{n;1}\xrightarrow{a\parallel b}\sum_{\lambda=\pm}\textrm{Split}^{\textrm{tree}}_{-\lambda}(a^{\lambda_a},b^{\lambda_b}) \mathcal{A}^{\textrm{loop}}_{n-1;1}(....(a+b)^{\lambda},...)\nonumber\\+\textrm{Split}^{\textrm{loop}}_{-\lambda}(a^-,b^-)\mathcal{A}^{\textrm{tree}}_{n-1}(....,(a+b)^{
\lambda},...).
\ee 
In the same helicity sector the second term in (\ref{coloop}) drops out because the split factor in this case multiples the same helicity tree amplitude that vanishes. Then, for the case of the all same helicity one loop amplitudes, the collinear limit takes the simple form 
\be 
\label{coloop2}
\mathcal{A}^{\textrm{one-loop}}_{n}(1^-,2^-,...,n^-)\xrightarrow{a\parallel b}\textrm{Split}^{\textrm{tree}}_-(a^-,b^-) \mathcal{A}^{\textrm{one-loop}}_{n-1}(....(a+b)^-,...),
\ee 
where the split function in this case is given by
\be 
\textrm{Split}_-^{\textrm{tree}}(a^-,b^-)=\frac{1}{\sqrt{z(1-z)}\langle ab\rangle} .
\ee 
Here $z$ is a parameter that describes how the two momenta $a,b$ become collinear, such that $a\rightarrow zP$ and $b\rightarrow (1-z)P$, where $P$ is null momentum.

\subsubsection{Verification of the correct collinear properties of the new formula}

Our expression (\ref{A-YM-general}) has the required collinear properties by construction. Indeed, we first note that in a colour-ordered amplitude only two adjacent momenta can contribute to the collinear singularity. Consider a pair $a,b$ that is colour-adjacent and goes collinear. In the sum in (\ref{A-YM-general}) this pair belongs to two types of terms. In the first type, both $a,b$ belong to the same Berends-Giele current. As we shall see in a moment, Berends-Giele currents have the correct collinear properties. So, these terms satisfy the collinear properties automatically.

The second type of terms is when $a,b$ belong to two groups of momenta that in (\ref{A-YM-general}) are separated by the bubble insertion. Let $a=j, b=j+1$. Let $p_j$ be the region momentum between $a,b$. Now, by region momentum independence of the amplitude (that should be established separately), we can set to zero any one of the region momenta. Let us choose to set to zero $p_j=0$. It is then easy to see that all the terms where $a,b$ do not belong to the same BG current get multiplied by $\langle q|p_j \circ p_i|q\rangle^2$, where $p_i$ is some other region momentum. Thus, with our choice $p_j=0$ all these terms do not contribute. All other terms have $a,b$ belong to the same BG current, i.e. all other terms are those of the first type. This shows that one can always write the one-loop amplitude in a form that exhibits the $a,b$ collinear property manifestly.

To complete the argument we can see that
\be
J(1, \ldots, j,j+1, \ldots, n) \rightarrow \frac{1}{\sqrt{z(1-z)}\langle ab\rangle} J(1,\ldots, P,\ldots, n),
\ee
so that the BG currents have the collinear property manifest. This follows from the formula (\ref{YM-BG}) and the fact that the momentum spinors for $a,b$ are given by
\be 
\label{rel} 
\lambda_a\rightarrow \sqrt{z}\lambda_P, ~~~~\lambda_b\rightarrow\sqrt{1-z}\lambda_P.
\ee
The only property that remains to be shown is the momentum region independence of the formula (\ref{A-YM-general}). We do this in the following subsection.

\subsection{Alternative ways of writing the formula}

Using the fact that 
\be
p_j = p_{i-1} + \sum_{l=i}^j k_l,
\ee
we can rewrite the formula (\ref{A-YM-general}) lowering the power of the region momenta in it
\be
{\cal A}= \sum_{part} J(i,\ldots, j) J(j+1,\ldots, i-1) \langle q| p_j \circ (\sum_{l=i}^j k_l) |q\rangle^2.
\ee

We can also rewrite the sum over partitions as a sum over cyclic permutations of the set $1,\ldots, n$. Indeed, it is easy to check that
\be\label{formula-cyclic}
2 {\cal A} = \sum_{i=1}^{n-1} J(1,\ldots, i) J(i+1,\ldots,n) \langle q| p_i \circ ( k_{i+1} + \ldots + k_n)| q\rangle^2 + {\rm cyclic},
\ee
where we need to add all cyclic permutations of the set $(1,\ldots, n)$. This form of the formula is particularly convenient for estabslishing the region momentum independence. Written in this way the amplitude formula is very similar to the one that appears in \cite{Skvortsov:2020gpn} in the light-cone gauge.

\subsection{Region momentum independence}

The purpose of this subsection is to argue that the amplitude is invariant under shifts of all region momenta by the same amount. Combined with our collinear limit argument, this gives a proof of the formula (\ref{A-YM-general}).

\subsubsection{Quadratic part of the dependence}

When we shift all region momenta by some value $x$ there are both quadratic and linear in $x$ terms that appear. Using (\ref{formula-cyclic}), the part quadratic in the shift  can be written as 
\be
{\cal A}^{x^2}= \sum_{i=1}^{n-1} J(1,\ldots, i) J(i+1,\ldots,n) \langle q| x \circ ( k_{i+1} + \ldots + k_n)| q\rangle^2 + {\rm cyclic}.
\ee
Because $x$ here is an arbitrary vector, so is the primed spinor $\langle q|x|:=\mu$. This means that we must consider
\be
\sum_{i=1}^{n-1} J(1,\ldots, i) J(i+1,\ldots,n) [\mu |  ( k_1 + \ldots + k_i)| q\rangle  [\mu | k_{i+1} + \ldots + k_{n}|q\rangle  + {\rm cyclic},
\ee
where we wrote the expression more symmetrically. This can be computed using the identity
\be
\sum_{i=1}^{n-1} J(1,\ldots, i) J(i+1,\ldots, n) [ \mu | k_1+ \ldots + k_i | q\rangle [\mu | k_{i+1} + \ldots + k_{n}|q\rangle = 
\frac{ [\mu | \sum_{i<j} i\circ j| \mu]}{\langle q1\rangle \langle 12\rangle \ldots \langle (n-1) n\rangle \langle nq\rangle},
\ee
which holds for arbitrary momenta $1,\ldots n$. The momenta in this formula are not assumed to add up to zero. This formula is proven analogously to how the recursive formula for the Berends-Giele currents is established. 

We will also need the identity
\be\label{identity}
\sum_{i=1}^{n-1} \frac{\langle i(i+1)\rangle}{\langle iq\rangle \langle (i+1)q\rangle} =\frac{\langle 1n\rangle}{\langle 1q\rangle \langle nq\rangle},
\ee
which is a simple consequence of Schouten identity. It can also be written as
\be
\sum_{i=1}^{n} \frac{\langle i(i+1)\rangle}{\langle iq\rangle \langle (i+1)q\rangle} =0,
\ee
with the convention that $n+1=1$. Using this identity we have
\be
\sum_{i=1}^{n-1} J(1,\ldots, i) J(i+1,\ldots,n) [\mu |  ( k_1 + \ldots + k_i)| q\rangle  [\mu | k_{i+1} + \ldots + k_{n}|q\rangle  + {\rm cyclic} = \\ \nonumber
\frac{2}{ \langle 12\rangle \ldots \langle (n-1)n\rangle \langle n1\rangle}
\left( \sum_{i<j} \frac{ [\mu| i\circ j|\mu] (ij)}{\langle iq\rangle \langle jq\rangle}\right).
\ee
No momentum conservation has yet been used. It is then easily checked that when the momentum conservation e.g. in the form $-n=1+\ldots + (n-1)$ is used, the coefficients in front of independent $[\mu|i\circ j|\mu]$ factors with $i,j = 1,\ldots, n-1$ vanish. Thus, the quadratic in $x$ part of dependence of the amplitude on the region momentum vanishes.

\subsubsection{Linear part of the dependence}

Taking the first variation of the amplitude as all region momenta vary, and denoting $\langle q| x|=[\mu|$ as before, we get a multiple of
\be
\sum_{i=1}^{n-1} J(1,\ldots, i) J(i+1,\ldots,n) [\mu| ( k_{1} + \ldots + k_i)| q\rangle 
\langle q| p_i \circ ( k_{i+1} + \ldots + k_n)| q\rangle
 + {\rm cyclic}.
\ee
The idea is again to compute the sum here explicitly, similar to what one does in the check of the Berends-Giele formula for the all same helicity currents. This is an exercise in applying Schouten identity. The result is
\be\label{proof-1}
\sum_{i=1}^{n-1} J(1,\ldots, i) J(i+1,\ldots,n) [\mu| ( k_1 + \ldots + k_i)| q\rangle 
\langle q| p_i \circ ( k_{i+1} + \ldots + k_n)| q\rangle = \\ \nonumber
\frac{1}{ \langle 1q\rangle \langle 12\rangle \ldots \langle (n-1) n\rangle \langle nq\rangle} 
\left( \sum_{i=1}^{n-1} \sum_{j>i} [\mu| i\circ j\circ p_i|q\rangle - \sum_{i=1}^{n-2} [\mu | i| q\rangle \sum_{j>i}^n \sum_{l>j}^n  s_{jl}  \right),
\ee
where the last terms contain Mandelstam variables $s_{ij}:= \langle ij\rangle [ij]$ and arise from relating the region momenta to each other via relations of the type $p_{i+1}=p_i + k_{i+1}$. 

It remains to add the cyclic permutations, and then apply the momentum conservation. Taking the cyclic permutation of the first set of terms in brackets in (\ref{proof-1}), and using (\ref{identity}) gives
\be
-\sum_{i=1}^{n-1} \sum_{j>i} \frac{\langle ij\rangle}{\langle iq\rangle \langle jq\rangle} ( [\mu|i\circ j\circ p_i|q\rangle - [\mu|j\circ i\circ p_j|q\rangle),
\ee
where from now on we omit the common prefactor
\be\label{prefactor}
\frac{1}{  \langle 12\rangle \ldots \langle (n-1) n\rangle \langle n1\rangle  } .
\ee
We now use the momentum conservation to express the last momentum $k_n$ in terms of all the rest. After this, we collect the terms in front of similar $[\mu| i\circ j\circ p|q\rangle$ expressions. Using (\ref{identity}) one more time we get for these terms
\be
- \sum_{i=1}^{n-1} \sum_{j=1}^{n-1} \frac{\langle in\rangle}{\langle iq\rangle \langle nq\rangle} [\mu| j\circ i\circ (p_n-p_j)|q\rangle. 
\ee
The sum here is taken over $i\not = j$. This is now written in terms of differences of region momenta, and so depends just on the external momenta $p_n-p_j =  -(k_1+ \ldots + k_j)$. We can thus rewrite the above as
\be\label{proof-2}
\sum_{i=1}^{n-1} \sum_{j=1, j\not= i}^{n-1}  \sum_{l=1}^j \frac{\langle in\rangle}{\langle iq\rangle \langle nq\rangle} [\mu| j\circ i\circ l| q\rangle. 
\ee

Let us now consider the quantity $ [\mu| j\circ i\circ l|q\rangle$
where $l$ is one of the momenta. We can exchange $i\circ l= - l \circ i - s_{il} \id$, where $s_{il}$ is the Mandelstam variable. On the other hand, we have
\be
\frac{\langle in\rangle}{\langle iq\rangle \langle nq\rangle} [\mu| j\circ l\circ i|q\rangle=
\frac{1}{\langle nq\rangle} [\mu j] \langle j l\rangle [l i] \langle in\rangle=
\frac{\langle jn\rangle}{\langle jq\rangle \langle nq\rangle} [\mu| j|q\rangle  \frac{\langle j|l\circ i|n\rangle}{\langle j n\rangle}.
\ee
This is linear in $i$, and so the sum over $i$ can be easily taken. Using the momentum conservation gives
\be
\sum_{i=1, i\not= j}^{n-1}\frac{\langle jn\rangle}{\langle jq\rangle \langle nq\rangle} [\mu| j|q\rangle  \frac{\langle j|l\circ i|n\rangle}{\langle j n\rangle}
= -\frac{\langle jn\rangle}{\langle jq\rangle \langle nq\rangle} [\mu| j|q\rangle  \frac{\langle j|l\circ (j+n)|n\rangle}{\langle j n\rangle} = \frac{\langle jn\rangle}{\langle jq\rangle \langle nq\rangle} [\mu| j|q\rangle s_{lj}.
\ee
Using these identities (\ref{proof-2}) becomes
\be\label{proof-3}
- \sum_{i=1}^{n-1} [\mu| i|q\rangle \sum_{j=1}^{n-1}  \frac{\langle jn\rangle}{\langle jq\rangle \langle nq\rangle}   \sum_{l=1}^i s_{jl} 
\ee

Let us now do the same manipulations with the second set of terms in (\ref{proof-1}). We add cyclic permutations, then use the momentum conservation to express $[\mu|n|q\rangle$ in terms of the other quantities of this type, use (\ref{identity}) as well as momentum conservation for the Mandelstam variables. We get precisely the terms in (\ref{proof-3}) with opposite signs, so these terms cancel each other and the linear part is zero. This finishes the proof.

\subsection{4-point amplitude}
The first non-trivial case is that of four gluons. For 3 gluons our formula produces an expression that can be rewritten in terms of Mandelstan variables, and these vanish at 3 points. 

In the case of 4 gluons, substituting the expressions for the currents we get 
\be\label{ym-4-point}
\mathcal{A}_4=\frac{1}{\langle q1\rangle \langle 1q\rangle} \frac{1}{\langle q2\rangle \langle 23\rangle \langle 34\rangle \langle 4q\rangle} \langle q| p_1\circ p_4|q\rangle^2+
\frac{1}{\langle q2\rangle \langle 2q\rangle} \frac{1}{\langle q3\rangle \langle 34\rangle \langle 41\rangle \langle 1q\rangle} \langle q| p_2\circ p_1|q\rangle^2+ \\ \nonumber
\frac{1}{\langle q3\rangle \langle 3q\rangle} \frac{1}{\langle q4\rangle \langle 41\rangle \langle 12\rangle \langle 2q\rangle} \langle q| p_3\circ p_2|q\rangle^2+
\frac{1}{\langle q4\rangle \langle 4q\rangle} \frac{1}{\langle q1\rangle \langle 12\rangle \langle 23\rangle \langle 3q\rangle} \langle q| p_4\circ p_3|q\rangle^2+ \\ \nonumber
\frac{1}{\langle q1\rangle \langle 12\rangle  \langle 2q\rangle} \frac{1}{\langle q3\rangle  \langle34\rangle \langle 4q\rangle} \langle q| p_4\circ p_2|q\rangle^2+
\frac{1}{\langle q2\rangle \langle 23\rangle \langle 3q\rangle} \frac{1}{\langle q4\rangle  \langle41\rangle \langle 1q\rangle} \langle q| p_3\circ p_1|q\rangle^2.
\ee
We know from general grounds that this expression must be region momentum and auxiliary spinor $q$ independent, and should match the known 4-point amplitude. But it is instructive to see how this happens.

We parametrise all region momenta in terms of one of them, e.g. $p_1=x$, and the external momenta. We have
\be
p_1=x, \quad p_2=2+x, \quad p_3=3+2 +x, \quad p_4=x-1.
\ee
We can then drop all terms containing $x$ because these terms vanish to render the result region momentum independent. This collapses the result to
\be
\mathcal{A}_4=\frac{1}{\langle q3\rangle \langle 3q\rangle} \frac{1}{\langle q4\rangle \langle 41\rangle \langle 12\rangle \langle 2q\rangle} \langle q| 3\circ 2|q\rangle^2+
\frac{1}{\langle q4\rangle \langle 4q\rangle} \frac{1}{\langle q1\rangle \langle 12\rangle \langle 23\rangle \langle 3q\rangle} \langle q| 1 \circ 4 |q\rangle^2+ \\ \nonumber
\frac{1}{\langle q1\rangle \langle 12\rangle  \langle 2q\rangle} \frac{1}{\langle q3\rangle  \langle34\rangle \langle 4q\rangle} \langle q| 1\circ 2|q\rangle^2 = 
 \frac{\langle 2q\rangle [32]^2}{\langle 4q\rangle  \langle 12\rangle \langle 41\rangle}+
 \frac{\langle 1q\rangle [14]^2}{ \langle 12\rangle \langle 23\rangle \langle 3q\rangle} +
 \frac{\langle 1q\rangle   \langle 2q\rangle [12]^2}{\langle 3q\rangle  \langle 12\rangle\langle34\rangle \langle 4q\rangle} .
\ee
We now eliminate $q$ dependence using the momentum conservation
\be
\frac{[32]}{\langle 41\rangle} = \frac{[12]}{\langle 34\rangle} , \qquad
\frac{[14]}{\langle 23\rangle} = \frac{[12]}{\langle 34\rangle} .
\ee
This gives
\be
\mathcal{A}_4= \frac{[12]}{\langle 12\rangle \langle 34\rangle \langle 3q\rangle\langle 4q\rangle} 
\left( \langle 3q\rangle \langle 2q\rangle [32] +\langle 1q\rangle \langle 4q\rangle [14] + \langle 1q\rangle   \langle 2q\rangle [12]\right)= \\\nonumber
\frac{[12][43]}{\langle 12\rangle \langle 34\rangle}
= \frac{[12][23]}{ \langle 34\rangle\langle 41\rangle} = 
\frac{s_{12}s_{23}}{ \langle 12\rangle\langle 23\rangle \langle 34\rangle\langle 41\rangle},
\ee
which is the correct answer (\ref{A4}) for this amplitude.

\subsection{Five point amplitude}

The expression for the amplitude in terms of currents is
\be
{\mathcal A}_5=J(1) J(2,3,4,5) \langle q| p_1\circ p_5 |q\rangle^2 + J(2) J(3,4,5,1) \langle q| p_2\circ p_1 |q\rangle^2 + J(3) J(4,5,1,2) \langle q| p_3\circ p_2 |q\rangle^2  \\ \nonumber
+J(4) J(5,1,2,3) \langle q| p_4\circ p_3 |q\rangle^2 + J(5) J(1,2,3,4) \langle q| p_5\circ p_4 |q\rangle^2  \\ \nonumber
+J(1,2) J(3,4,5)  \langle q| p_2\circ p_5 |q\rangle^2 + J(2,3) J(4,5,1)  \langle q| p_3\circ p_1 |q\rangle^2
+J(3,4) J(5,1,2)  \langle q| p_4\circ p_2 |q\rangle^2 \\ \nonumber
+J(4,5) J(1,2,3)  \langle q| p_5\circ p_3 |q\rangle^2+J(5,1) J(2,3,4)  \langle q| p_1\circ p_4 |q\rangle^2.
\ee
Again, we know that it must reproduce the known answer, but would like to see explicitly how this happens. This requires much more work as compared to the 4-point case.

\subsubsection{Extracting the region momentum independent result}

We again parametrise the region momenta in terms of one of them, and the external momenta
\be
p_1=x, \quad p_2=2+x, \quad p_3=3+2 +x, \quad p_4=4+3+2+x, \quad p_5=x-1.
\ee
All terms containing $x$ must drop out by region momentum independence. This gives the following expression
\be
 {\mathcal A}_5=J(3) J(4,5,1,2) \langle q| 3\circ 2 |q\rangle^2 + 
J(4) J(5,1,2,3) \langle q| 4\circ (3+2) |q\rangle^2 + J(5) J(1,2,3,4) \langle q| 1\circ 5 |q\rangle^2  \\ \nonumber
+J(1,2) J(3,4,5)  \langle q| 2\circ 1 |q\rangle^2 
+J(3,4) J(5,1,2)  \langle q| (4+3)\circ 2 |q\rangle^2+J(4,5) J(1,2,3)  \langle q| (4+5)\circ 1 |q\rangle^2.
\ee
Substituting the expressions for the currents we get
\be
 {\mathcal A}_5=\frac{[23]^2\langle 2q\rangle}{ \langle 4q\rangle \langle 45\rangle \langle 51\rangle \langle 12\rangle} + \frac{([41]\langle 1q\rangle + [45]\langle 5q\rangle)^2}{ \langle 5q\rangle \langle 51\rangle \langle 12\rangle \langle 23\rangle \langle 3q\rangle}
 +\frac{[15]^2\langle 1q\rangle}{ \langle 12\rangle \langle 23\rangle \langle 34\rangle \langle 4q\rangle}
 \\ \nonumber
 + \frac{[12]^2\langle 1q\rangle\langle 2q\rangle}{ \langle 3q\rangle  \langle 12\rangle \langle 34\rangle \langle 45\rangle \langle 5q\rangle}
 + \frac{([21]\langle 1q\rangle + [25]\langle 5q\rangle)^2 \langle 2q\rangle}{ \langle 3q\rangle \langle 4q\rangle \langle 5q\rangle\langle 34\rangle \langle 51\rangle \langle 12\rangle}
 + \frac{([14]\langle 4q\rangle + [15]\langle 5q\rangle)^2 \langle 1q\rangle}{\langle 3q\rangle  \langle 4q\rangle \langle 5q\rangle \langle 45\rangle \langle 12\rangle \langle 23\rangle}.
 \ee
 Let us start by bringing it all to the common denominator
 \be\nonumber
  {\mathcal A}_5= \frac{1}{\langle 3q\rangle  \langle 4q\rangle \langle 5q\rangle \langle 12\rangle  \langle 23\rangle  \langle 34\rangle \langle 45\rangle  \langle 51\rangle}\times\\ \nonumber
  \Big(
  [23]^2\langle 2q\rangle\langle 3q\rangle\langle 5q\rangle\langle 23\rangle  \langle 34\rangle
  + [15]^2\langle 1q\rangle \langle 3q\rangle\langle 5q\rangle  \langle 45\rangle  \langle 51\rangle
  + [12]^2\langle 1q\rangle\langle 2q\rangle\langle 4q\rangle \langle 23\rangle\langle 51\rangle
  \\ \nonumber
+([41]\langle 1q\rangle + [45]\langle 5q\rangle)^2 \langle 4q\rangle \langle 34\rangle \langle 45\rangle
   + ([23]\langle 3q\rangle + [24]\langle 4q\rangle)^2 \langle 2q\rangle \langle 23\rangle \langle 45\rangle
  +([14]\langle 4q\rangle + [15]\langle 5q\rangle)^2 \langle 1q\rangle \langle 34\rangle \langle 51\rangle
  \Big).
  \ee
 We then expand the squares and collect the terms next to common square bracket factors. One then notices that such terms can be rewritten  more compactly using Schouten identity
 \be
 [15]^2\langle 1q\rangle \langle 3q\rangle\langle 5q\rangle  \langle 45\rangle  \langle 51\rangle 
 +  [15]^2 \langle 5q\rangle^2 \langle 1q\rangle \langle 34\rangle \langle 51\rangle
 = [15]^2 \langle 1q\rangle \langle 4q\rangle \langle 5q\rangle \langle 51\rangle \langle 35\rangle
 \\ \nonumber
 [23]^2\langle 2q\rangle\langle 3q\rangle\langle 5q\rangle\langle 23\rangle  \langle 34\rangle
 + [23]^2\langle 3q\rangle^2  \langle 2q\rangle \langle 23\rangle \langle 45\rangle=
 [23]^2\langle 2q\rangle  \langle 3q\rangle  \langle 4q\rangle \langle 23\rangle \langle 35\rangle
 \\ \nonumber
 [14]^2\langle 1q\rangle^2 \langle 4q\rangle \langle 34\rangle \langle 45\rangle+
 [14]^2 \langle 4q\rangle^2 \langle 1q\rangle \langle 34\rangle \langle 51\rangle
 =[14]^2\langle 1q\rangle \langle 4q\rangle \langle 5q\rangle\langle 34\rangle \langle 41\rangle . 
 \ee
 This gives for the amplitude
 \be\label{A5-result-q}
   {\mathcal A}_5= \frac{1}{\langle 3q\rangle  \langle 4q\rangle \langle 5q\rangle \langle 12\rangle  \langle 23\rangle  \langle 34\rangle \langle 45\rangle  \langle 51\rangle}\times\\ \nonumber
  \Big(
 s_{23} [23]\langle 2q\rangle  \langle 3q\rangle  \langle 4q\rangle  \langle 35\rangle
  + s_{15} [15] \langle 1q\rangle \langle 4q\rangle \langle 5q\rangle  \langle 53\rangle
  + [12]^2\langle 1q\rangle\langle 2q\rangle\langle 4q\rangle \langle 23\rangle\langle 51\rangle
  \\ \nonumber
+ s_{45} [45]\langle 5q\rangle^2 \langle 4q\rangle \langle 34\rangle 
   +  [24]^2 \langle 4q\rangle^2 \langle 2q\rangle \langle 23\rangle \langle 45\rangle
  +s_{14} [14] \langle 1q\rangle \langle 4q\rangle \langle 5q\rangle\langle 43\rangle 
    \\ \nonumber 
    2 s_{45} [41]\langle 1q\rangle  \langle 5q\rangle \langle 4q\rangle \langle 34\rangle 
   + 2 s_{23} \langle 3q\rangle  [24]\langle 4q\rangle \langle 2q\rangle  \langle 45\rangle
  +2 s_{15} [14]\langle 4q\rangle \langle 5q\rangle \langle 1q\rangle \langle 43\rangle 
  \Big),
  \ee
  where we used the notation $s_{ij} := \langle ij\rangle [ij]$.
  
  To understand the steps that follow we start by writing the amplitude that we want to reproduce. Our starting point is the form (\ref{A5-altern}) of the amplitude. It is clear that in this expression there are terms that can be written in terms of Mandelstam variables, but there is always a remainder that cannot be written in this way. In the formula (\ref{A5-altern}) this is the last term. This term, however, can be written in many different ways. Let us first massage it into the form that will be useful later.

We use the momentum conservation in the form $-\langle 23\rangle [34] = \langle 21\rangle [14] +\langle 25\rangle [54]$ to rewrite
\be
\langle 12\rangle  \langle 23\rangle  \langle 34\rangle \langle 45\rangle  \langle 51\rangle  {\mathcal A}_5 =  s_{12} s_{23}  +s_{45} s_{51} + s_{25}s_{45}+ \langle 21\rangle [14] \langle 45\rangle [52].
\ee
Finally, we use Schouten identity in the last term to rewrite the amplitude as
 \be\label{A5-full}
\langle 12\rangle  \langle 23\rangle  \langle 34\rangle \langle 45\rangle  \langle 51\rangle  {\mathcal A}_5 =  s_{12} s_{23}  +s_{45} s_{51} + s_{25}s_{45}+ s_{25} s_{14} +\langle 24\rangle [14] \langle 15\rangle [52].
\ee

The idea now is to see which of the terms in the amplitude (\ref{A5-result-q}) can reproduce the last term in (\ref{A5-full}). Most of the terms in (\ref{A5-result-q})  already contain factors of Mandelstam variables, and so cannot be responsible for this term. The only terms that can be responsible are the ones containing $[12]^2$ and $[24]^2$. To massage these terms into the desired form we use 
\be
- [21] \langle 1q\rangle  = [23] \langle 3q\rangle  + [24] \langle 4q\rangle + [25] \langle 5q\rangle.
\ee
This gives, using the momentum conservation in terms proportional to $[24]$
\be
[12]^2\langle 1q\rangle\langle 2q\rangle\langle 4q\rangle \langle 23\rangle\langle 51\rangle
+ [24]^2 \langle 4q\rangle^2 \langle 2q\rangle \langle 23\rangle \langle 45\rangle = \\ \nonumber
s_{23} [21] \langle 2q\rangle \langle 3q\rangle \langle 4q\rangle    \langle 15\rangle
+[21] [25] \langle 2q\rangle  \langle 4q\rangle \langle 5q\rangle \langle 23\rangle  \langle 15\rangle
- s_{23} [24] \langle 4q\rangle^2 \langle 2q\rangle \langle 35\rangle .
\ee
We then use for the middle term
\be\nonumber
[21]\langle 23\rangle = [14] \langle 43\rangle + [15] \langle 53\rangle, 
\ee 
to get
\be\nonumber
[12]^2\langle 1q\rangle\langle 2q\rangle\langle 4q\rangle \langle 23\rangle\langle 51\rangle
+ [24]^2 \langle 4q\rangle^2 \langle 2q\rangle \langle 23\rangle \langle 45\rangle = \\ \nonumber
s_{23} [21] \langle 2q\rangle \langle 3q\rangle \langle 4q\rangle    \langle 15\rangle
- s_{23} [24] \langle 4q\rangle^2 \langle 2q\rangle \langle 35\rangle
+ [25]  [14] \langle 43\rangle  \langle 2q\rangle  \langle 4q\rangle \langle 5q\rangle \langle 15\rangle + s_{15} [25]  \langle 53\rangle \langle 2q\rangle  \langle 4q\rangle \langle 5q\rangle .
\ee
As the last step, we extract the $q$-independent part of the third term using Schouten identity. 
\be\label{A5-interm-1}
[12]^2\langle 1q\rangle\langle 2q\rangle\langle 4q\rangle \langle 23\rangle\langle 51\rangle
+ [24]^2 \langle 4q\rangle^2 \langle 2q\rangle \langle 23\rangle \langle 45\rangle = 
[52]  [14] \langle 24\rangle \langle 15\rangle \langle 3q\rangle  \langle 4q\rangle \langle 5q\rangle 
\\ \nonumber
+s_{23} [21] \langle 2q\rangle \langle 3q\rangle \langle 4q\rangle    \langle 15\rangle
- s_{23} [24] \langle 4q\rangle^2 \langle 2q\rangle \langle 35\rangle
 + s_{15} [25]  \langle 53\rangle \langle 2q\rangle  \langle 4q\rangle \langle 5q\rangle 
 \\ \nonumber
 + [25]  [14] \langle 23\rangle    \langle 4q\rangle^2 \langle 5q\rangle \langle 15\rangle.
\ee
The term on the right-hand side of the first line (after dividing by the $q$-dependent terms in the denominator) is precisely the last term in (\ref{A5-full}) that can not be written in terms of Mandelstam variables. The term on the last line can also be written in terms of Mandelstam variables. Indeed, we first use Schouten identity $ \langle 15\rangle \langle 4q\rangle =\langle 14\rangle \langle 5q\rangle-\langle 1q\rangle \langle 54\rangle$ to write
\be
[25]  [14] \langle 23\rangle    \langle 4q\rangle^2 \langle 5q\rangle \langle 15\rangle=
s_{14} [25]  \langle 23\rangle    \langle 4q\rangle \langle 5q\rangle^2 - [25]  [14] \langle 23\rangle    \langle 1q\rangle \langle 4q\rangle \langle 5q\rangle \langle 54\rangle.
\ee
We then use $- [25]\langle 54\rangle = [21]\langle 14\rangle+[23]\langle 34\rangle$ to finally get
\be
[25]  [14] \langle 23\rangle    \langle 4q\rangle^2 \langle 5q\rangle \langle 15\rangle=
s_{14} [25]  \langle 23\rangle    \langle 4q\rangle \langle 5q\rangle^2 
+ s_{14} [21] \langle 23\rangle \langle 1q\rangle \langle 4q\rangle \langle 5q\rangle 
+ s_{23} [14] \langle 34\rangle \langle 1q\rangle \langle 4q\rangle \langle 5q\rangle.
\ee

It thus remains to reproduce the other $s$-containing terms in the formula (\ref{A5-full}) for the amplitude. We substitute the terms in the second and third line of (\ref{A5-interm-1}) into (\ref{A5-result-q}) instead of the $[12]^2, [24]^2$ terms. This gives a part of the amplitude that is supposed to contain all terms with Mandelstam variables
\be
 \langle 3q\rangle  \langle 4q\rangle \langle 5q\rangle \langle 12\rangle  \langle 23\rangle  \langle 34\rangle \langle 45\rangle  \langle 51\rangle   {\mathcal A}'_5 = s_{23} [23] \langle 35\rangle \langle 2q\rangle  \langle 3q\rangle  \langle 4q\rangle  
  + s_{15} [15] \langle 53\rangle \langle 1q\rangle \langle 4q\rangle \langle 5q\rangle 
  \\ \nonumber
+ s_{45} [45]\langle 34\rangle   \langle 4q\rangle \langle 5q\rangle^2
  +s_{14} [14] \langle 43\rangle \langle 1q\rangle \langle 4q\rangle \langle 5q\rangle
    \\ \nonumber 
 +   2 s_{45} [14] \langle 43\rangle \langle 1q\rangle  \langle 4q\rangle \langle 5q\rangle 
   + 2 s_{23}   [24] \langle 45\rangle \langle 2q\rangle \langle 3q\rangle \langle 4q\rangle 
  +2 s_{15} [14] \langle 43\rangle\langle 1q\rangle \langle 4q\rangle \langle 5q\rangle  
 \\ \nonumber
+s_{23} [21] \langle 15\rangle\langle 2q\rangle \langle 3q\rangle \langle 4q\rangle    
- s_{23} [24] \langle 35\rangle \langle 2q\rangle\langle 4q\rangle^2  
 + s_{15} [25]  \langle 53\rangle \langle 2q\rangle  \langle 4q\rangle \langle 5q\rangle 
 \\ \nonumber
 + s_{14} [25]  \langle 23\rangle    \langle 4q\rangle \langle 5q\rangle^2 
+ s_{14} [21] \langle 23\rangle \langle 1q\rangle \langle 4q\rangle \langle 5q\rangle 
- s_{23} [14] \langle 43\rangle \langle 1q\rangle \langle 4q\rangle \langle 5q\rangle.
\ee
There are some immediate simplifications. The terms containing $s_{23}  \langle 2q\rangle  \langle 3q\rangle  \langle 4q\rangle $ simplify using $[23]\langle 35\rangle + [21]\langle 15\rangle=- [24]\langle 45\rangle$. The terms containing $[14] \langle 43\rangle \langle 1q\rangle \langle 4q\rangle \langle 5q\rangle$ simplify using $s_{14}+2s_{45} + 2s_{15} -s_{23}=s_{23}-s_{14}$, and so
\be
 \langle 3q\rangle  \langle 4q\rangle \langle 5q\rangle \langle 12\rangle  \langle 23\rangle  \langle 34\rangle \langle 45\rangle  \langle 51\rangle   {\mathcal A}'_5 = 
    s_{23}   [24] \langle 45\rangle \langle 2q\rangle \langle 3q\rangle \langle 4q\rangle    
   \\ \nonumber
  + s_{15} [15] \langle 53\rangle \langle 1q\rangle \langle 4q\rangle \langle 5q\rangle 
+ s_{45} [45]\langle 34\rangle   \langle 4q\rangle \langle 5q\rangle^2  
 \\ \nonumber
- s_{23} [24] \langle 35\rangle \langle 2q\rangle\langle 4q\rangle^2  
 + s_{15} [25]  \langle 53\rangle \langle 2q\rangle  \langle 4q\rangle \langle 5q\rangle 
 \\ \nonumber
 + s_{14} [25]  \langle 23\rangle    \langle 4q\rangle \langle 5q\rangle^2 
+ s_{14} [21] \langle 23\rangle \langle 1q\rangle \langle 4q\rangle \langle 5q\rangle 
+ (s_{23} -s_{14})[14] \langle 43\rangle \langle 1q\rangle \langle 4q\rangle \langle 5q\rangle.
\ee
The two terms in the last line containing $s_{14} \langle 1q\rangle \langle 4q\rangle \langle 5q\rangle$ simplify using $-[12]\langle 23\rangle - [14]\langle 43\rangle = [15]\langle 53\rangle$, and so
\be
 \langle 3q\rangle  \langle 4q\rangle \langle 5q\rangle \langle 12\rangle  \langle 23\rangle  \langle 34\rangle \langle 45\rangle  \langle 51\rangle   {\mathcal A}'_5 = 
    s_{23}   [24] \langle 45\rangle \langle 2q\rangle \langle 3q\rangle \langle 4q\rangle    
   \\ \nonumber
  + (s_{15}+s_{14})  [15] \langle 53\rangle \langle 1q\rangle \langle 4q\rangle \langle 5q\rangle 
+ s_{45} [45]\langle 34\rangle   \langle 4q\rangle \langle 5q\rangle^2  
 \\ \nonumber
- s_{23} [24] \langle 35\rangle \langle 2q\rangle\langle 4q\rangle^2  
 + s_{15} [25]  \langle 53\rangle \langle 2q\rangle  \langle 4q\rangle \langle 5q\rangle 
 \\ \nonumber
 + s_{14} [25]  \langle 23\rangle    \langle 4q\rangle \langle 5q\rangle^2 
+ s_{23} [14] \langle 43\rangle \langle 1q\rangle \langle 4q\rangle \langle 5q\rangle.
\ee
We then again use the same momentum conservation formula on the very last term to get
\be
 \langle 3q\rangle  \langle 4q\rangle \langle 5q\rangle \langle 12\rangle  \langle 23\rangle  \langle 34\rangle \langle 45\rangle  \langle 51\rangle   {\mathcal A}'_5 = 
    s_{23}   [24] \langle 45\rangle \langle 2q\rangle \langle 3q\rangle \langle 4q\rangle    
   \\ \nonumber
  -s_{45}  [15] \langle 53\rangle \langle 1q\rangle \langle 4q\rangle \langle 5q\rangle 
+ s_{45} [45]\langle 34\rangle   \langle 4q\rangle \langle 5q\rangle^2  
 \\ \nonumber
- s_{23} [24] \langle 35\rangle \langle 2q\rangle\langle 4q\rangle^2  
 + s_{15} [25]  \langle 53\rangle \langle 2q\rangle  \langle 4q\rangle \langle 5q\rangle 
 \\ \nonumber
 + s_{14} [25]  \langle 23\rangle    \langle 4q\rangle \langle 5q\rangle^2 
- s_{23} [12] \langle 23\rangle \langle 1q\rangle \langle 4q\rangle \langle 5q\rangle.
\ee
We can now use Schouten identity to extract the $q$-invariant pieces, and match these to those in (\ref{A5-full}). We have
\be
- s_{23} [12] \langle 23\rangle \langle 1q\rangle \langle 4q\rangle \langle 5q\rangle=
s_{23} s_{12} \langle 3q\rangle \langle 4q\rangle \langle 5q\rangle +
s_{23} [12] \langle 31\rangle \langle 2q\rangle \langle 4q\rangle \langle 5q\rangle, 
\\ \nonumber
-s_{45}  [15] \langle 53\rangle \langle 1q\rangle \langle 4q\rangle \langle 5q\rangle =
s_{45} s_{51} \langle 3q\rangle \langle 4q\rangle \langle 5q\rangle +
s_{45} [15] \langle 31\rangle  \langle 4q\rangle \langle 5q\rangle^2, 
\\ \nonumber
 s_{14} [25]  \langle 23\rangle    \langle 4q\rangle \langle 5q\rangle^2 =
 s_{14} s_{25} \langle 3q\rangle \langle 4q\rangle \langle 5q\rangle +
s_{14} [25] \langle 53\rangle \langle 2q\rangle \langle 4q\rangle \langle 5q\rangle.
\ee
These gives three of the four Mandelstam variable containing terms in (\ref{A5-full}). The remainder, which is supposed to give the last $s_{25} s_{45}$ term is 
\be
  s_{23}   [24] \langle 45\rangle \langle 2q\rangle \langle 3q\rangle \langle 4q\rangle 
  +s_{45} [15] \langle 31\rangle  \langle 4q\rangle \langle 5q\rangle^2
  + s_{45} [45]\langle 34\rangle   \langle 4q\rangle \langle 5q\rangle^2  
 \\ \nonumber
- s_{23} [24] \langle 35\rangle \langle 2q\rangle\langle 4q\rangle^2  
 + s_{15} [25]  \langle 53\rangle \langle 2q\rangle  \langle 4q\rangle \langle 5q\rangle 
 \\ \nonumber
 +s_{14} [25] \langle 53\rangle \langle 2q\rangle \langle 4q\rangle \langle 5q\rangle
 +s_{23} [12] \langle 31\rangle \langle 2q\rangle \langle 4q\rangle \langle 5q\rangle.
 \ee
The second and third terms here, using the momentum conservation, give
\be
s_{45} [25] \langle 23\rangle  \langle 4q\rangle \langle 5q\rangle^2 =
s_{45} s_{25} \langle 3q\rangle \langle 4q\rangle \langle 5q\rangle +
s_{45} [25] \langle 53\rangle \langle 2q\rangle \langle 4q\rangle \langle 5q\rangle.
\ee
The first term gives the last term in (\ref{A5-full}). Thus, we have the remainder which is
\be
  s_{23}   [24] \langle 45\rangle \langle 2q\rangle \langle 3q\rangle \langle 4q\rangle 
  +s_{45} [25] \langle 53\rangle \langle 2q\rangle \langle 4q\rangle \langle 5q\rangle
 \\ \nonumber
- s_{23} [24] \langle 35\rangle \langle 2q\rangle\langle 4q\rangle^2  
 + s_{15} [25]  \langle 53\rangle \langle 2q\rangle  \langle 4q\rangle \langle 5q\rangle 
 \\ \nonumber
 +s_{14} [25] \langle 53\rangle \langle 2q\rangle \langle 4q\rangle \langle 5q\rangle
 +s_{23} [12] \langle 31\rangle \langle 2q\rangle \langle 4q\rangle \langle 5q\rangle.
 \ee
 Applying Schouten identity once more on the first term in the first and second lines
 \be
s_{23}   [24] \langle 45\rangle \langle 2q\rangle \langle 3q\rangle \langle 4q\rangle
-s_{23} [24] \langle 35\rangle \langle 2q\rangle\langle 4q\rangle^2
=s_{23}   [24] \langle 43\rangle \langle 2q\rangle \langle 4q\rangle \langle 5q\rangle,
\ee
we get a set of terms all proportional to $\langle 2q\rangle \langle 4q\rangle \langle 5q\rangle$
\be\nonumber
s_{23}   [24] \langle 43\rangle+s_{23} [21] \langle 13\rangle + s_{45} [25] \langle 53\rangle+s_{15} [25] \langle 53\rangle +s_{14} [25] \langle 53\rangle 
=[25] \langle 53\rangle ( s_{45} + s_{15} + s_{14} - s_{23})=0,
\ee
where we applied momentum conservation to the first two terms. Thus, the correct expression (\ref{A5-full}) for the 5-point amplitude is reproduced. 

\section{Discussion}

This paper establishes a new way of understanding the same helicity one-loop Yang-Mills amplitudes as arising from tree-level Berends-Giele currents connected by an effective propagator. This new interpretation is important for several reasons. 

The so-called CSW \cite{Cachazo:2004kj} tree-level formalism gives a prescription for how to compute tree-level YM amplitudes with effective Feynman rules that use MHV amplitudes as vertices. An attempt to extend this to loop amplitudes was made in \cite{Cachazo:2004zb}, where it was in particular suggested that one-loop same helicity amplitudes that are the subject of this paper can be added as new interaction vertices, in addition to the tree level MHV vertices. The interpretation of these one-loop amplitudes that emerges from our work is that they are build from more elementary tree-level blocks. It thus seems unnatural to try to build more involved amplitudes from the amplitudes that are already composed. Our results show that these amplitudes are made from simpler tree-level amplitudes connected by an effective propagator. Thus, it appears that it is the region momenta-dependent effective propagator that should be added into the set of CSW rules if one is to reproduce loop amplitudes. 

Our new interpretation is also relevant for the problem of UV divergence of quantum gravity. In papers \cite{Bern:2015xsa}, \cite{Bern:2017puu} this divergence was linked to the non-vanishing of the same helicity one-loop amplitude in GR. Given that the story with GR is likely mirroring that in YM, see more on this below, the interpretation of this paper shows that the non-vanishing of the same helicity one-loop amplitudes can in turn be linked to the non-vanishing of the one-loop bubble. It would thus be very interesting to better understand the significance and interpretation of the non-vanishing of the bubbles. 

This leads to the question of interpretation of such a simple construction as our formula (\ref{A-YM-general}). This formula cries for an interpretation other than just an observation that the formula works. The fact that region momentum variables play such a prominent role points to both the worldsheet interpretation as in \cite{Chakrabarti:2005ny}, as well as the (momentum) twistor space interpretation as in \cite{Mason:2009qx}. In any case, it would be very interesting to understand a deeper origin of the formula (\ref{A-YM-general}). 

Another natural question is how much of what was described here can be generalised to the case of gravity. As in the case of YM, there exists a truncation of the full theory of General Relativity that keeps only its self-dual sector. In its version that most parallels the story of SDYM, this has been described in \cite{Krasnov:2016emc}. The corresponding "flat" version relevant for GR with zero scalar curvature has been recently described in \cite{Krasnov:2021cva}. The "flat" version that is most relevant for computing flat space amplitudes has properties very similar to that of SDYM. There is a single kinetic term, and then a single cubic interaction, which however in the gravity case contains two derivatives. The Feynman rules for this SDGR theory are immediate, and can be used for computing amplitudes. In particular, the all-negative BG currents can be computed by a recursion, see e.g. \cite{Krasnov:2016emc} and references therein for the description of the result and the procedure. The integrand of the one-loop 4 point amplitude can also be written down without any difficulty. 

What can also be studied without difficulty is the "bubble" self-energy diagram. Analogously to the SDYM case, this amplitude can be argued to be zero, while at the same time shift dependent. The shift can be computed, with the result being the most natural generalisation of the effective propagator (\ref{eff-prop})  from spin one two spin two case. What is far from clear, however, is how to interpret the shift parameters in the gravity case. In the case of YM we could associate these with the region momenta. There is no cyclic ordering in the case of gravity, and so the use of region momenta does not seem justified any longer. Thus, while there is strong evidence that there must exist a gravitational analog of the formula (\ref{A-YM-general}), it is not clear how to to give meaning to the momentum variables appearing in the gravitational effective propagator, and so it is not clear what form the GR analog of (\ref{A-YM-general}) should take. We hope to return to all these questions in a future publication.

\section*{Acknowledgements} We are grateful to Eugene Skvortsov for pointing out the "sum over bubbles" interpretation of the one-loop amplitudes, and in particular for the reference \cite{Chakrabarti:2005ny}. 

\appendix

\section{Sum of box, triangle and bubble diagrams}

The purpose of this Appendix is to spell out, in our notations, the argument that the sum of the box, triangle and bubble one-loop integrands, with a specific choice of the loop momentum, equivalent to the parametrisation in terms of the region momenta, vanishes. This means that the 4-point one-loop amplitude can be reduced to "bubble" insertions. Our formula (\ref{A-YM-general}) is then the generalisation of this statement to an arbitrary number of external states. 

\subsection{Box Diagram}

The box diagram 
\vspace{1em}
\be
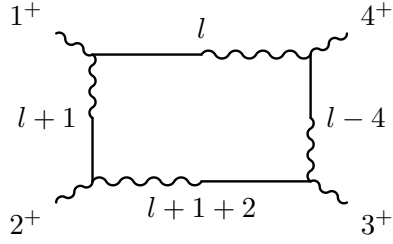
\begin{fmfgraph*}(110,80)
\fmfbottom{i1,i2}
\fmftop{o1,o2}
\fmf{photon,tension=3}{i1,v1}
\fmf{photon,tension=3}{i2,v2}
\fmf{photon,tension=3}{o1,v3}
\fmf{photon,tension=3}{o2,v4}
\fmf{photon}{v3,v3'}
\fmf{vanilla}{v3',v1}
\fmf{photon}{v1,v1'}
\fmf{vanilla}{v1',v2}
\fmf{photon}{v2,v2'}
\fmf{vanilla}{v2',v4}
\fmf{photon}{v4,v4'}
\fmf{vanilla}{v4',v3}
\fmflabel{$1^+$}{o1}
\fmflabel{$4^+$}{o2}
\fmflabel{$2^+$}{i1}
\fmflabel{$3^+$}{i2}
\fmflabel{$l+1+2$}{v1'}
\fmflabel{$l+1$}{v3'}
\fmflabel{$l-4$}{v2'}
\fmflabel{$l$}{v4'}
\end{fmfgraph*}\nonumber
\ee
has already been studied by us in \cite{Chattopadhyay:2020oxe}. It is given by
\be 
\label{amp}
i\mathcal{M}=\int \frac{d^4l}{(2\pi)^4}\frac{\langle q|l|4]\langle q|l+1|1]\langle q|l+1+2|2]\langle q|l-4|3]}{l^2(l+1)^2(l+1+2)^2(l-4)^2\prod_{j=1}^4\langle qj\rangle}.
\ee
We use manipulations described in \cite{Chattopadhyay:2020oxe} to reduce it to a sum of triangle-like and then bubble-like diagrams, using various identities to cancel the scalar propagator factors of the type $l^2$ from the denominator. The first stage of reduction is described by
\be 
2i\mathcal{M}=2i\mathcal{M}_1+2i\mathcal{M}_2+2i\mathcal{M},
_3
\ee 
with
\be
\begin{split}
\label{ammpli}
2i\mathcal{M}_1&=\int \frac{d^4l}{(2\pi)^4}\frac{\langle q|3\circ(l-4)|q\rangle\langle q|l+1|1]\langle q|l+1+2|2]}{(l+1)^2(l+1+2)^2(l-4)^2\prod_{j=1}^4\langle qj\rangle\langle 43\rangle}\\\nonumber
2i\mathcal{M}_2&=\int \frac{d^4l}{(2\pi)^4}\frac{\langle q|l\circ (l-4)|q\rangle\langle q|l+1|1]\langle q|l+1+2|2]}{l^2(l+1)^2(l-4)^2\prod_{j=1}^4\langle qj\rangle\langle 43\rangle}\\\nonumber
2i\mathcal{M}_3&=\int \frac{d^4l}{(2\pi)^4}\frac{\langle q|l\circ(l-3-4)|q\rangle\langle q|l+1|1]\langle q|l+1+2|2]}{l^2(l+1)^2(l+1+2)^2\prod_{j=1}^4\langle qj\rangle\langle 43\rangle}.
\end{split}
\ee
In the second stage of reduction one gets, see Section 3 of \cite{Chattopadhyay:2020oxe} for derivation
\be 
\label{amp111}
4i\mathcal{M}_{1}=\int \frac{d^4l}{(2\pi)^4}\frac{\langle q|l+1|1]\langle q|(l-4)\circ(1+4)|q\rangle}{(l+1)^2(l-4)^2\langle 23\rangle\langle 43\rangle\langle q1\rangle\langle q2\rangle\langle q4\rangle}
\\\nonumber
+\int \frac{d^4l}{(2\pi)^4}\frac{\langle q|l+1|1]\langle q|(l+1+2)\circ 3|q\rangle}{(l+1+2)^2(l-4)^2\langle 23\rangle\langle 43\rangle\langle q1\rangle\langle q2\rangle\langle q4\rangle}
\\\nonumber
+\int \frac{d^4l}{(2\pi)^4}\frac{\langle q|l+1|1]\langle q|(l+1+2)\circ2|q\rangle}{(l+1)^2(l+1+2)^2\langle 23\rangle\langle 43\rangle\langle q1\rangle\langle q2\rangle\langle q4\rangle}
\ee
\be 
\label{amp232}
4i\mathcal{M}_{2}=\int \frac{d^4l}{(2\pi)^4}\frac{\langle q|l+1|2]\langle q|l\circ1|q\rangle}{l^2(l+1)^2\langle 41\rangle\langle 43\rangle\langle q1\rangle\langle q2\rangle\langle q3\rangle}\\\nonumber
+\int \frac{d^4l}{(2\pi)^4}\frac{\langle q|l+1|2]\langle q|(l-4)\circ(2+3)|q\rangle}{(l+1)^2(l-4)^2\langle 41\rangle\langle 43\rangle\langle q1\rangle\langle q2\rangle\langle q3\rangle}\\\nonumber+\int \frac{d^4l}{(2\pi)^4}\frac{\langle q|l+1|2]\langle q|l\circ4|q\rangle}{l^2(l-4)^2\langle 41\rangle\langle 43\rangle\langle q1\rangle\langle q2\rangle\langle q3\rangle}
\ee
\be 
\label{amp332}
4i\mathcal{M}_{3}=\int \frac{d^4l}{(2\pi)^4}\frac{\langle q|l\circ(3+4)|q\rangle\langle q|(1+2)\circ l|q\rangle}{l^2(l+1+2)^2\langle 12\rangle\langle 43\rangle\langle q1\rangle\langle q2\rangle\langle q3\rangle\langle q4\rangle}\\\nonumber
+\int \frac{d^4l}{(2\pi)^4}\frac{\langle q|l\circ(3+4)|q\rangle\langle q|(l+1)\circ 1|q\rangle}{l^2(l+1)^2\langle 12\rangle\langle 43\rangle\langle q1\rangle\langle q2\rangle\langle q3\rangle\langle q4\rangle}\\
+\nonumber\int \frac{d^4l}{(2\pi)^4}\frac{\langle q|l\circ(3+4)|q\rangle\langle q|(l+1)\circ2|q\rangle}{(l+1)^2(l+1+2)^2\langle 12\rangle\langle 43\rangle\langle q1\rangle\langle q2\rangle\langle q3\rangle\langle q4\rangle}
\ee

We now implement the use of region momenta $p_i$, which we assign as follows
\be 
k_1=p_1-p_4, \quad
k_2=p_2-p_1, \quad
k_3=p_3-p_2, \quad
k_4=p_4-p_3.
\ee 
We also assign the region momentum for the the loop interior and re-express the loop momentum in terms of region momenta. Since the loop momentum has to be integrated, there is an arbitrariness in its assignment. Indeed, we can add any momenta $q$ to the loop. We thus make a specific choice $l=x-p_4$, where $x$ is the region momentum inside the loop and $p_4$ is one of the external region momenta. The momentum $x$ is now to be integrated over. We then re-write the quadratically divergent integrals in terms of the dual momentum variables and usual momenta, given this particular choice of the loop momentum. This gives
\be 
4i\mathcal{M}_1=4i\mathcal{M}_{11}+4i\mathcal{M}_{12}+4i\mathcal{M}_{13},
\ee 
where
\be
\begin{split}
\label{amp111}
4i\mathcal{M}_{11}&=\int \frac{d^4x}{(2\pi)^4}\frac{\langle q|p_1-x|1]\langle q|(p_2-x)\circ2|q\rangle}{(p_2-x)^2(p_1-x)^2\langle 23\rangle\langle 43\rangle\langle q1\rangle\langle q2\rangle\langle q4\rangle}\\\nonumber
4i\mathcal{M}_{12}&=\int \frac{d^4x}{(2\pi)^4}\frac{\langle q|p_1-x|1]\langle q|(p_3-x)\circ(1+4)|q\rangle}{(p_1-x)^2(p_3-x)^2\langle 23\rangle\langle 43\rangle\langle q1\rangle\langle q2\rangle\langle q4\rangle}\\\nonumber
4i\mathcal{M}_{13}&=\int \frac{d^4x}{(2\pi)^4}\frac{\langle q|p_1-x|1]\langle q|(p_2-x)\circ3|q\rangle}{(p_3-x)^2(p_2-x)^2\langle 23\rangle\langle 43\rangle\langle q1\rangle\langle q2\rangle\langle q4\rangle},
\end{split}
\ee
\be 
4i\mathcal{M}_2=4i\mathcal{M}_{21}+4i\mathcal{M}_{22}+4i\mathcal{M}_{23},
\ee 
where
\be 
\begin{split}
\label{amp2321}
4i\mathcal{M}_{21}&=\int \frac{d^4x}{(2\pi)^4}\frac{\langle q|p_1-x|2]\langle q|(p_4-x)\circ1|q\rangle}{(p_4-x)^2(p_1-x)^2\langle 41\rangle\langle 43\rangle\langle q1\rangle\langle q2\rangle\langle q3\rangle}\\\nonumber
4i\mathcal{M}_{22}&=\int \frac{d^4x}{(2\pi)^4}\frac{\langle q|p_1-x|2]\langle q|(p_3-x)\circ(3+2)|q\rangle}{(p_1-x)^2(p_3-x)^2\langle 41\rangle\langle 43\rangle\langle q1\rangle\langle q2\rangle\langle q3\rangle}\\\nonumber
4i\mathcal{M}_{23}&=\int \frac{d^4x}{(2\pi)^4}\frac{\langle q|p_1-x|2]\langle q|(p_4-x)\circ4|q\rangle}{(p_4-x)^2(p_3-x)^2\langle 41\rangle\langle 43\rangle\langle q1\rangle\langle q2\rangle\langle q3\rangle},
\end{split}
\ee
and finally
\be 
4i\mathcal{M}_3=4i\mathcal{M}_{31}+4i\mathcal{M}_{32}+4i\mathcal{M}_{33},
\ee 
where
\be 
\begin{split}
\label{amp3321}
4i\mathcal{M}_{31}&=\int \frac{d^4x}{(2\pi)^4}\frac{\langle q|(p_4-x)\circ(3+4)|q\rangle\langle q|(p_2-x)\circ (3+4)|q\rangle}{(p_4-x)^2(p_2-x)^2\langle 12\rangle\langle 43\rangle\langle q1\rangle\langle q2\rangle\langle q3\rangle\langle q4\rangle}\\
4i\mathcal{M}_{32}&=\int \frac{d^4x}{(2\pi)^4}\frac{\langle q|(p_4-x)\circ(3+4)|q\rangle\langle q|(p_1-x)\circ 1|q\rangle}{(p_4-x)^2(p_1-x)^2\langle 12\rangle\langle 43\rangle\langle q1\rangle\langle q2\rangle\langle q3\rangle\langle q4\rangle}\\
4i\mathcal{M}_{33}&=\int \frac{d^4x}{(2\pi)^4}\frac{\langle q|(p_4-x)\circ(3+4)|q\rangle\langle q|(p_2-x)\circ2|q\rangle}{(p_1-x)^2(p_2-x)^2\langle 12\rangle\langle 43\rangle\langle q1\rangle\langle q2\rangle\langle q3\rangle\langle q4\rangle}
\end{split}
\ee

\subsection{Triangle Diagrams}
There are four in-equivalent triangle diagrams to consider. We start by considering the one where the legs $3$ and $4$ are attached to two of the vertices.
\vspace{1em}
\vspace{1em}

~~~~~~~~~~~~~~~~~~~~~~~~~~~~~~~~~~~~~~~~~
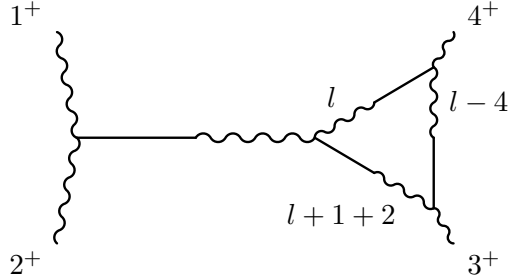
\begin{fmfgraph*}(150,100)
\fmfbottom{i1,i2}
\fmftop{o1,o2}
\fmf{photon,tension=3}{i1,v1,o1}
\fmf{photon,tension=3}{i2,v3}
\fmf{photon,tension=3}{o2,v4}
\fmf{vanilla}{v1,v1'}
\fmf{photon}{v1',v2}
\fmf{photon,label=$l$}{v2,v2'}
\fmf{vanilla}{v2',v4}
\fmf{photon,label=$l-4$}{v4',v4}
\fmf{vanilla}{v4',v3}
\fmf{photon,label=$l+1+2$}{v3',v3}
\fmf{vanilla}{v3',v2}
\fmflabel{$1^+$}{o1}
\fmflabel{$4^+$}{o2}
\fmflabel{$2^+$}{i1}
\fmflabel{$3^+$}{i2}
\fmflabel{}{v1'}
\fmflabel{}{v2'}
\fmflabel{}{v3'}
\fmflabel{}{v4'}
\end{fmfgraph*}\\~\\
\be 
2i\mathcal{T}^{12}=\frac{1}{\langle 12\rangle}\int \frac{d^4l}{(2\pi)^4}\frac{\langle q|l|4]\langle q|l-4|3]\langle q|(l+1+2)\circ(1+2)|q\rangle}{l^2(l-4)^2(l+1+2)^2\prod_{j=1}^4\langle qj\rangle}
\ee 
Let us start by multiplying the numerator and denominator by $\langle 34\rangle$. This allows to write the integral as 
\be 
2i\mathcal{T}^{12}=-\frac{1}{\langle 12\rangle\langle 34\rangle}\int \frac{d^4l}{(2\pi)^4}\frac{\langle q|l\circ 4\circ 3\circ(l-4)|q\rangle\langle q|l\circ(1+2)|q\rangle}{l^2(l-4)^2(l+1+2)^2\prod_{j=1}^4\langle qj\rangle},
\ee 
where we used $\langle q|(1+2)\circ(1+2)|q\rangle=0$ to get rid of $(1+2)$ from the last factor in the numerator. We now replace $4=l-(l-4)$ and use the identity $l\circ l=\frac{1}{2}l^2\mathds{1}$ to write the integral as
\be 
2i\mathcal{T}^{12}=-\frac{1}{\langle 12\rangle\langle 34\rangle}\Bigg[-\frac{1}{2}\int \frac{d^4l}{(2\pi)^4}\frac{\langle q|(l-4)\circ 3|q\rangle\langle q|l\circ(1+2)|q\rangle}{(l-4)^2(l+1+2)^2\prod_{j=1}^4\langle qj\rangle}
\nonumber\\-\int \frac{d^4l}{(2\pi)^4}\frac{\langle q|l\circ(l-4)\circ 3\circ(l-4)|q\rangle\langle q|l\circ(1+2)|q\rangle}{l^2(l-4)^2(l+1+2)^2\prod_{j=1}^4\langle qj\rangle}\Bigg]
\ee 
Next we use the spinor identity 
\be 
A\circ B=-B\circ A+(A.B)\mathds{1}
\ee 
where $(A.B)$ is the metric pairing. This identity holds for any two arbitrary rank two mixed spinors $A$ and $B$. Using this, we have 
\be 
\label{id}
(l-4)\circ 3\circ(l-4)=-(l-4)\circ(l-4)\circ3+(l-4)((l-4).3)\nonumber\\=-\frac{1}{2}(l-4)^2 3+(l-4)((l-4).3)
\ee 
The second term in the last line of (\ref{id}) can be written as 
\be 
(l-4).3=\frac{1}{2}((l-4)^2-(l-4-3)^2)
\ee 
Therefore, this gives us 
\be 
(l-4)\circ 3\circ(l-4)=\frac{1}{2}(l-4)^2 (l-4-3)-\frac{1}{2}(l-4-3)^2(l-4)
\ee 
We now replace $l-3-4=l+1+2$ and cancel denominators to get 
\be 
2i\mathcal{T}^{12}=\frac{1}{\langle 12\rangle\langle 34\rangle}\Bigg[\frac{1}{2}\int \frac{d^4l}{(2\pi)^4}\frac{\langle q|(l-4)\circ 3|q\rangle\langle q|l\circ(1+2)|q\rangle}{(l-4)^2(l+1+2)^2\prod_{j=1}^4\langle qj\rangle}
\nonumber\\
+\frac{1}{2}\int \frac{d^4l}{(2\pi)^4}\frac{\langle q|l\circ(l+1+2)|q\rangle\langle q|l\circ(1+2)|q\rangle}{l^2(l+1+2)^2\prod_{j=1}^4\langle qj\rangle}
\nonumber\\-\frac{1}{2}\int \frac{d^4l}{(2\pi)^4}\frac{\langle q|l\circ(l-4)|q\rangle\langle q|l\circ(1+2)|q\rangle}{l^2(l-4)^2\prod_{j=1}^4\langle qj\rangle}\Bigg]
\ee 
One of the factors of $l$ in the second and third integrals can be immediately eliminated owing to the identity $\langle q|l\circ l|q\rangle=0$. We then choose to rewrite it as 
\be 
\label{tri}
i\mathcal{T}^{12}=\frac{1}{4\langle 12\rangle\langle 34\rangle}\Bigg[\int \frac{d^4l}{(2\pi)^4}\frac{\langle q|(l-4)\circ 3|q\rangle\langle q|(l+1+2)\circ(1+2)|q\rangle}{(l-4)^2(l+1+2)^2\prod_{j=1}^4\langle qj\rangle}
\nonumber\\
+\int \frac{d^4l}{(2\pi)^4}\frac{\langle q|l\circ(1+2)|q\rangle\langle q|(l+1+2)\circ(1+2)|q\rangle}{l^2(l+1+2)^2\prod_{j=1}^4\langle qj\rangle}
\nonumber\\
+\int \frac{d^4l}{(2\pi)^4}\frac{\langle q|l\circ4|q\rangle\langle q|(l+1+2)\circ(1+2)|q\rangle}{l^2(l-4)^2\prod_{j=1}^4\langle qj\rangle}\Bigg]
\ee 
Let us then rewrite this in the dual momentum variables. As before, we choose the loop momentum variable $l=p_4-x$. We have  
\be 
4i\mathcal{T}^{12}=4i\mathcal{T}^{12}_1+4i\mathcal{T}^{12}_2+4i\mathcal{T}^{12}_3
\ee 
where 
\be 
\begin{split}
\label{tri2}
4i\mathcal{T}^{12}_1&=\int \frac{d^4l}{(2\pi)^4}\frac{\langle q|(p_3-x)\circ 3|q\rangle\langle q|(p_2-x)\circ(1+2)|q\rangle}{(p_3-x)^2(p_2-x)^2\langle 12\rangle\langle 34\rangle\prod_{j=1}^4\langle qj\rangle}\\
4i\mathcal{T}^{12}_2&=\int \frac{d^4l}{(2\pi)^4}\frac{\langle q|(p_4-x)\circ(1+2)|q\rangle\langle q|(p_2-x)\circ(1+2)|q\rangle}{(p_4-x)^2(p_2-x)^2\langle 12\rangle\langle 34\rangle\prod_{j=1}^4\langle qj\rangle}\\
4i\mathcal{T}^{12}_3&=\int \frac{d^4l}{(2\pi)^4}\frac{\langle q|(p_4-x)\circ4|q\rangle\langle q|(p_2-x)\circ(1+2)|q\rangle}{(p_4-x)^2(p_3-x)^2\langle 12\rangle\langle 34\rangle\prod_{j=1}^4\langle qj\rangle}
\end{split}
\ee 

The diagram with legs 1 and 2 inserted to two of the vertices is
\vspace{1em}

~~~~~~~~~~~~~~~~~~~~~~~~~~~~
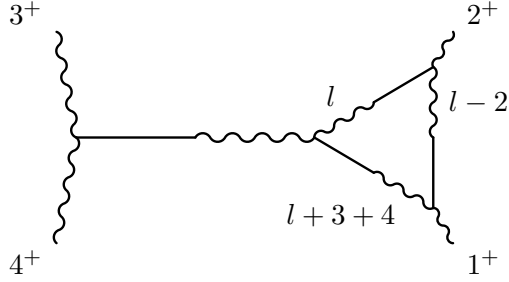
\begin{fmfgraph*}(150,100)
\fmfbottom{i1,i2}
\fmftop{o1,o2}
\fmf{photon,tension=3}{i1,v1,o1}
\fmf{photon,tension=3}{i2,v3}
\fmf{photon,tension=3}{o2,v4}
\fmf{vanilla}{v1,v1'}
\fmf{photon}{v1',v2}
\fmf{photon,label=$l$}{v2,v2'}
\fmf{vanilla}{v2',v4}
\fmf{photon,label=$l-2$}{v4',v4}
\fmf{vanilla}{v4',v3}
\fmf{photon,label=$l+3+4$}{v3',v3}
\fmf{vanilla}{v3',v2}
\fmflabel{$3^+$}{o1}
\fmflabel{$2^+$}{o2}
\fmflabel{$4^+$}{i1}
\fmflabel{$1^+$}{i2}
\fmflabel{}{v1'}
\fmflabel{}{v2'}
\fmflabel{}{v3'}
\fmflabel{}{v4'}
\end{fmfgraph*}
\\~\\
It can be obtained from (\ref{tri2}) by substitutions
\be
1\to 3, \quad 2\to 4,\quad 3\to 1,\quad 4\to 2.
\ee
The loop momentum in terms of region momenta is $l=p_2-x$. We have
\be 
4i\mathcal{T}^{34}=4i\mathcal{T}^{34}_1+4i\mathcal{T}^{34}_2+4i\mathcal{T}^{34}_3,
\ee 
where 
\be 
\begin{split}
\label{tri31}
4i\mathcal{T}^{34}_1&=\int \frac{d^4x}{(2\pi)^4}\frac{\langle q|(p_1-x)\circ 1|q\rangle\langle q|(p_4-x)\circ (3+4)|q\rangle}{(p_1-x)^2(p_4-x)^2\langle 12\rangle\langle 34\rangle\langle q1\rangle\langle q2\rangle\langle q3\rangle\langle q4\rangle}\\
4i\mathcal{T}^{34}_2&=\int \frac{d^4x}{(2\pi)^4}\frac{\langle q|(p_2-x)\circ(3+4)|q\rangle\langle q|(p_4-x)\circ (3+4)|q\rangle}{(p_2-x)^2(p_4-x)^2\langle 12\rangle\langle 34\rangle\langle q1\rangle\langle q2\rangle\langle q3\rangle\langle q4\rangle}\\
4i\mathcal{T}^{34}_3&=\int \frac{d^4x}{(2\pi)^4}\frac{\langle q|(p_2-x)\circ 2|q\rangle\langle q|(p_4-x)\circ(3+4)|q\rangle}{(p_2-x)^2(p_1-x)^2\langle 12\rangle\langle 34\rangle\langle q1\rangle\langle q2\rangle\langle q3\rangle\langle q4\rangle}
\end{split}
\ee 
We note that the terms arising here cancel precisely the terms in ${\mathcal M}_3$. 

The diagram with legs 2 and 3 inserted to two of the vertices is

\vspace{1em}
~~~~~~~~~~~~~~~~~~~~~~~~~~~~~~~~~~
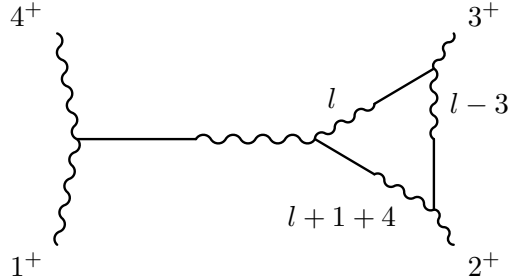
\begin{fmfgraph*}(150,100)
\fmfbottom{i1,i2}
\fmftop{o1,o2}
\fmf{photon,tension=3}{i1,v1,o1}
\fmf{photon,tension=3}{i2,v3}
\fmf{photon,tension=3}{o2,v4}
\fmf{vanilla}{v1,v1'}
\fmf{photon}{v1',v2}
\fmf{photon,label=$l$}{v2,v2'}
\fmf{vanilla}{v2',v4}
\fmf{photon,label=$l-3$}{v4',v4}
\fmf{vanilla}{v4',v3}
\fmf{photon,label=$l+1+4$}{v3',v3}
\fmf{vanilla}{v3',v2}
\fmflabel{$4^+$}{o1}
\fmflabel{$3^+$}{o2}
\fmflabel{$1^+$}{i1}
\fmflabel{$2^+$}{i2}
\fmflabel{}{v1'}
\fmflabel{}{v2'}
\fmflabel{}{v3'}
\fmflabel{}{v4'}
\end{fmfgraph*}
\\~\\
It can be obtained from (\ref{tri2})  by substitutions
\be
1\to 4, \quad 4\to 3,\quad 3\to 2,\quad 2\to 1.
\ee
The loop momentum in terms of region momenta is $l=p_3-x$. We get
\be 
4i\mathcal{T}^{41}=4i\mathcal{T}^{41}_1+4i\mathcal{T}^{41}_2+4i\mathcal{T}^{41}_3,
\ee 
where 
\be 
\begin{split}
\label{tri3}
4i\mathcal{T}^{41}_1&=\int \frac{d^4x}{(2\pi)^4}\frac{\langle q|(p_2-x)\circ 2|q\rangle\langle q|(p_1-x)\circ(1+4)|q\rangle}{(p_2-x)^2(p_1-x)^2\langle 41\rangle\langle 23\rangle\prod_{j=1}^4\langle qj\rangle}\\
4i\mathcal{T}^{41}_2&=\int \frac{d^4x}{(2\pi)^4}\frac{\langle q|(p_3-x)\circ(1+4)|q\rangle\langle q|(p_1-x)\circ(1+4)|q\rangle}{(p_3-x)^2(p_1-x)^2\langle 41\rangle\langle 23\rangle\prod_{j=1}^4\langle qj\rangle}\\
4i\mathcal{T}^{41}_3&=\int \frac{d^4x}{(2\pi)^4}\frac{\langle q|(p_3-x)\circ3|q\rangle\langle q|(p_1-x)\circ(1+4)|q\rangle}{(p_2-x)^2(p_3-x)^2\langle 41\rangle\langle 23\rangle\prod_{j=1}^4\langle qj\rangle}
\end{split}
\ee

The diagram with legs 1 and 4 inserted to two of the vertices is
\vspace{1em}

~~~~~~~~~~~~~~~~~~~~~~~~~~~~~~~
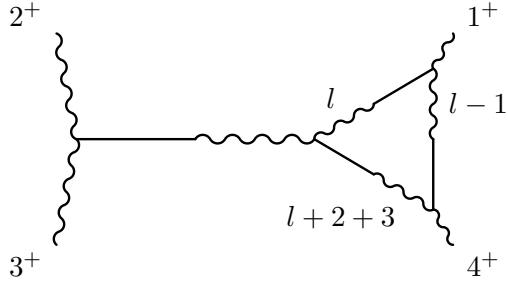
\begin{fmfgraph*}(150,100)
\fmfbottom{i1,i2}
\fmftop{o1,o2}
\fmf{photon,tension=3}{i1,v1,o1}
\fmf{photon,tension=3}{i2,v3}
\fmf{photon,tension=3}{o2,v4}
\fmf{vanilla}{v1,v1'}
\fmf{photon}{v1',v2}
\fmf{photon,label=$l$}{v2,v2'}
\fmf{vanilla}{v2',v4}
\fmf{photon,label=$l-1$}{v4',v4}
\fmf{vanilla}{v4',v3}
\fmf{photon,label=$l+2+3$}{v3',v3}
\fmf{vanilla}{v3',v2}
\fmflabel{$2^+$}{o1}
\fmflabel{$1^+$}{o2}
\fmflabel{$3^+$}{i1}
\fmflabel{$4^+$}{i2}
\fmflabel{}{v1'}
\fmflabel{}{v2'}
\fmflabel{}{v3'}
\fmflabel{}{v4'}
\end{fmfgraph*}
\\~\\
It can be obtained from (\ref{tri2}) by substitutions
\be
1\to 2, \quad 2\to 3,\quad 3\to 4,\quad 4\to 1.
\ee
The loop momentum in terms of region momenta is $l=p_1-x$. This gives
\be 
4i\mathcal{T}^{23}=4i\mathcal{T}^{23}_1+4i\mathcal{T}^{23}_2+4i\mathcal{T}^{23}_3,
\ee 
where 
\be 
\begin{split}
\label{tri31}
4i\mathcal{T}^{23}_1&=\int \frac{d^4x}{(2\pi)^4}\frac{\langle q|(p_4-x)\circ 4|q\rangle\langle q|(p_3-x)\circ(2+3)|q\rangle}{(p_4-x)^2(p_3-x)^2\langle 23\rangle\langle 41\rangle\prod_{j=1}^4\langle qj\rangle}\\
4i\mathcal{T}^{23}_2&=\int \frac{d^4x}{(2\pi)^4}\frac{\langle q|(p_1-x)\circ (2+3)|q\rangle\langle q|(p_3-x)\circ(2+3)|q\rangle}{(p_1-x)^2(p_3-x)^2\langle 23\rangle\langle 41\rangle\prod_{j=1}^4\langle qj\rangle}\\
4i\mathcal{T}^{23}_3&=\int \frac{d^4x}{(2\pi)^4}\frac{\langle q|(p_1-x)\circ 1|q\rangle\langle q|(p_3-x)\circ(2+3)|q\rangle}{(p_1-x)^2(p_4-x)^2\langle 23\rangle\langle 41\rangle\prod_{j=1}^4\langle qj\rangle}.
\end{split}
\ee

\subsection{Bubble Diagrams}

\subsubsection{Internal Bubbles}

There are two inequivalent permutations of the bubbles on internal lines. In one diagram, the legs 1 and 2 join on one side of the internal line while 3 and 4 on the other side. This is given by  \\~\\

~~~~~~~~~~~~~~~~~~~~~~~~~~~~~~~~~
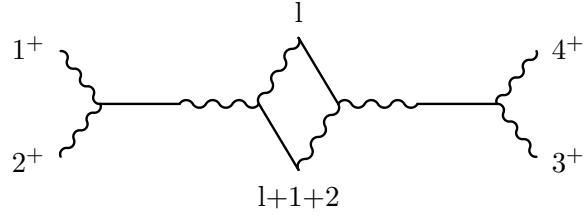
\begin{fmfgraph*}(180,50)
     \fmftop{i1,i3,i2}
     \fmfbottom{o1,o3,o2}
     \fmf{photon}{i1,v2,o1}
     \fmf{vanilla}{v2,v1'}
     \fmf{photon}{v1',v1}
     \fmf{photon}{v1,i3}
     \fmf{vanilla}{i3,v4}
     \fmf{photon}{v4,o3}
     \fmf{vanilla}{o3,v1}
     \fmf{photon}{v4,v3}
     \fmf{vanilla}{v3,v3'}
     \fmf{photon}{i2,v3',o2}
     \fmflabel{l}{i3}
     \fmflabel{l+1+2}{o3}
     \fmflabel{$1^+$}{i1}
     \fmflabel{$2^+$}{o1}
     \fmflabel{$3^+$}{o2}
     \fmflabel{$4^+$}{i2}
\end{fmfgraph*}
\\~\\

 \be 
4i\mathcal{B}^{12}=\frac{1}{\langle 12\rangle\langle 34\rangle}\int \frac{d^4l}{(2\pi)^4}\frac{\langle q|l\circ (1+2)|q\rangle\langle q|(l+1+2)\circ(3+4)|q\rangle}{l^2(l+1+2)^2\prod_{j=1}^4\langle qj\rangle}
\ee 
Writing it in the dual momentum variables with $l=p_4-x$ we get
 \be 
4i\mathcal{B}^{12}=\frac{1}{\langle 12\rangle\langle 34\rangle}\int \frac{d^4x}{(2\pi)^4}\frac{\langle q|(p_4-x)\circ (1+2)|q\rangle\langle q|(p_2-x)\circ(3+4)|q\rangle}{(p_4-x)^2(p_2-x)^2\prod_{j=1}^4\langle qj\rangle}.
\ee 
The other internal bubble diagram is
\\~\\

~~~~~~~~~~~~~~~~~~~~~~~~~~~~~~~~~
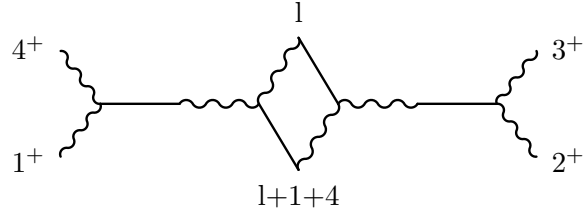
\begin{fmfgraph*}(180,50)
     \fmftop{i1,i3,i2}
     \fmfbottom{o1,o3,o2}
     \fmf{photon}{i1,v2,o1}
     \fmf{vanilla}{v2,v1'}
     \fmf{photon}{v1',v1}
     \fmf{photon}{v1,i3}
     \fmf{vanilla}{i3,v4}
     \fmf{photon}{v4,o3}
     \fmf{vanilla}{o3,v1}
     \fmf{photon}{v4,v3}
     \fmf{vanilla}{v3,v3'}
     \fmf{photon}{i2,v3',o2}
     \fmflabel{l}{i3}
     \fmflabel{l+1+4}{o3}
     \fmflabel{$4^+$}{i1}
     \fmflabel{$1^+$}{o1}
     \fmflabel{$2^+$}{o2}
     \fmflabel{$3^+$}{i2}
\end{fmfgraph*}
\\~\\
It can be obtained from the above by permutations, with $l=p_3-x$ and the result being
\be 
4i\mathcal{B}^{23}=\frac{1}{\langle 41\rangle\langle 23\rangle}\int \frac{d^4x}{(2\pi)^4}\frac{\langle q|(p_3-x)\circ (1+4)|q\rangle\langle q|(p_1-x)\circ(3+2)|q\rangle}{(p_3-x)^2(p_1-x)^2\prod_{j=1}^4\langle qj\rangle}.
\ee 

\subsubsection{Bubbles on External Lines}

There are four distinct diagrams here, corresponding to the different ways of insertion of bubbles to four external lines. One can write each contribution as an insertion of a current to one of the legs of the bubble. For example, we have for the insertion of the bubble between the particle 1 and the $J(2,3,4)$ current
\\~\\ 
\be
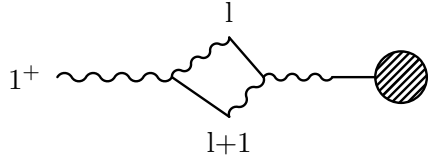
\begin{gathered}
~~~\begin{fmfgraph*}(130,30)
     \fmfleft{i}
     \fmfright{o}
     \fmfbottom{v}
     \fmftop{u}
     \fmf{photon}{i,v1}
     \fmf{photon}{v1,u}
     \fmf{vanilla}{u,v2}
     \fmf{photon}{v2,v}
     \fmf{vanilla}{v,v1}
     \fmf{photon}{v2,v3}
     \fmf{vanilla}{v3,o}
     \fmfblob{.15w}{o}
     \fmflabel{l}{u}
     \fmflabel{l+1}{v}
     \fmflabel{$1^+$}{i}
\end{fmfgraph*}
\end{gathered}\nonumber 
\ee 
\\
\be 
4i\mathcal{B}^1=\frac{\langle q3\rangle}{\langle 23\rangle\langle 34\rangle}\int \frac{d^4l}{(2\pi)^4}\frac{\langle q|(l+1)\circ 1|q\rangle\langle q|l|1]}{l^2(l+1)^2\prod_{j=1}^4\langle qj\rangle}
\ee 
Rewriting this in terms of the dual momentum variables with $l=p_4-x$ we get
\be 
4i\mathcal{B}^1=\frac{\langle q3\rangle}{\langle 23\rangle\langle 34\rangle}\int \frac{d^4x}{(2\pi)^4}\frac{\langle q|(p_1-x)\circ 1|q\rangle\langle q|p_4-x|1]}{(p_4-x)^2(p_1-x)^2\prod_{j=1}^4\langle qj\rangle}.
\ee 

The other similar contributions are obtained by cyclic permutations. We have
\be 
4i\mathcal{B}^2=\frac{\langle q4\rangle}{\langle 34\rangle\langle 41\rangle}\int \frac{d^4x}{(2\pi)^4}\frac{\langle q|(p_2-x)\circ 2|q\rangle\langle q|p_1-x|2]}{(p_2-x)^2(p_1-x)^2\prod_{j=1}^4\langle qj\rangle},
\ee 
\be 
4i\mathcal{B}^3=\frac{\langle q1\rangle}{\langle 41\rangle\langle 12\rangle}\int \frac{d^4x}{(2\pi)^4}\frac{\langle q|(p_3-x)\circ 3|q\rangle\langle q|p_2-x|3]}{(p_2-x)^2(p_3-x)^2\prod_{j=1}^4\langle qj\rangle},
\ee 
\be 
4i\mathcal{B}^4=\frac{\langle q2\rangle}{\langle 12\rangle\langle 23\rangle}\int \frac{d^4x}{(2\pi)^4}\frac{\langle q|(p_4-x)\circ 4|q\rangle\langle q|p_3-x|4]}{(p_4-x)^2(p_3-x)^2\prod_{j=1}^4\langle qj\rangle}
\ee

\subsection{Sum of Integrands}

As we previously noticed, the terms in ${\mathcal M}_3$ get cancelled by one of the triangle diagrams ${\mathcal T}^{34}$. Our aim is now to show that all other terms get cancelled agains each other as well. To this end, we will group the terms according to their denominators of the form 
$(p_i-x)^2(p_j-x)^2$ for some $i,j$. 

We start with the denominator factor $(p_1-x)^2(p_2-x)^2$. The contributions to this come from ${\mathcal M}_{11}, {\mathcal T}^{41}_1$ as well as one of the bubbles ${\mathcal B}^2$. 
The sum of the corresponding numerators is given by 
\be 
\label{n1}
\begin{split} 
&\frac{\langle q|(p_2-x)\circ 2|q\rangle}{\langle 23\rangle\langle 43\rangle \langle 14\rangle}\Big(\langle q|(p_1-x)\circ (1+4)|4\rangle\langle q3\rangle
-\langle q|(p_1-x)\circ (1+4)|q\rangle\langle 43\rangle
\\&~~~~~~~~~~~~~~~~~~~~~~~~~~
+\langle q|(p_1-x)\circ 2|3\rangle\langle q4\rangle\Big),
\end{split}
\ee 
where we rewrote the first term in a suggestive way.
We now use the momentum conservation $2=-(1+3+4)$ in the last term, and then Schouten identity 
$|4\rangle \langle q3\rangle = |q\rangle \langle 43\rangle + |3\rangle \langle q4\rangle$ to see that the sum in brackets is zero. 

Next consider the denominator factor $(p_1-x)^2(p_3-x)^2$. The contributions come from $\mathcal{M}_{12},\mathcal{M}_{22}, {\mathcal T}^{41}_2, {\mathcal T}^{23}_2$ and the bubble 
$\mathcal{B}^{23}$. The triangle contributions double each other, and the sum of these numerators is 
\be 
\label{n2}
\begin{split} 
&\frac{\langle q|(p_3-x)\circ(1+4)|q\rangle}{\langle 23\rangle\langle 43\rangle \langle 14\rangle}\Big(\langle q|(p_1-x)\circ (1+4) |4\rangle\langle q3\rangle+\langle q|(p_1-x)\circ (2+3)| 3\rangle\langle q4\rangle\\&
-2\langle q|(p_1-x)\circ (1+4)|q\rangle\langle 43\rangle
+\langle q|(p_1-x)\circ (1+4)|q\rangle\langle 43\rangle\Big),
\end{split}
\ee 
where again we rewrote the first terms in a suggestive way. Relacing $2+3=-(1+4)$ in the second term and using the same Schouten identity as above we see the cancellation.

Next consider the denominator factor $(p_1-x)^2(p_4-x)^2$. The contributions come from $\mathcal{M}_{21},\mathcal{T}^{23}_3$ and ${\mathcal B}^1$. The sum of these numerators is given by 
\be 
\label{n3}
\begin{split} 
&\frac{\langle q|(p_4-x)\circ1|q\rangle}{\langle 23\rangle\langle 43\rangle \langle 41\rangle}\Big(
\langle q|(p_1-x)\circ 2|3\rangle\langle q4\rangle
+\langle q|(p_3-1)\circ (2+3)|q\rangle\langle 43\rangle
\\&~~~~~~~~~~~~~~~~~~~~~~~~~~
+\langle q|(p_4-x)\circ 1|4\rangle\langle q3\rangle\Big).
\end{split}
\ee 
We now use the fact that $p_3=p_1+2+3$, and so $p_3$ can be replaced by $p_1$ in the second term. Similarly, $p_4=p_1-1$, and so we can replace $p_4$ with $p_1$ in the last term. We then similarly replace $2$ by $(2+3)$ in the first term, and $1$ by $(1+4)$ in the last. Then again the same Schouten identity implies the cancellation. 

Let us now consider the denominator factor $(p_2-x)^2(p_3-x)^2$. The situation is somewhat more interesting here. There are 4 integrands that contribute, namely ${\mathcal M}_{13}, {\mathcal T}^{12}_1, {\mathcal T}^{41}_3$ and the bubble ${\mathcal B}^3$. Let us start by considering the sum of ${\mathcal T}^{41}_3$ and ${\mathcal B}^3$. This can be written as
\be
\frac{\langle q| (p_3-x)\circ 3|q\rangle}{ \langle 41\rangle \langle 12\rangle \langle 23\rangle}
\left( \langle q|(p_1-x)\circ (1+4)|q\rangle \langle 12\rangle - \langle q|(p_2-x)\circ 3|2\rangle\langle q1\rangle \right).
\ee
Using $p_3-p_1=3+2$ in the first term, as well as $1+4=-(2+3)$, we can replace $p_1$ there by $p_3$. Similarly, in the second term we can use $p_3-p_2=3$ to replace $p_2$ with $p_3$. The expression in brackets is then
\be
-\langle q|(p_3-x)\circ (2+3)|q\rangle \langle 12\rangle - \langle q|(p_3-x)\circ (2+3)|2\rangle\langle q1\rangle = -\langle q|(p_3-x)\circ (2+3)|1\rangle \langle q2\rangle 
\ee
by Schouten identity. This can be written as
\be\label{app-1}
-\langle q|(p_3-x)\circ (2+3)|1\rangle \langle q2\rangle = -\langle q|(p_3-x)\circ (1+2+3)|1\rangle \langle q2\rangle = \langle q|(p_3-x)|4] \langle 41\rangle \langle q2\rangle.
\ee

On the other hand, the sum of ${\mathcal M}_{13}$ and ${\mathcal T}^{12}_1$ is given by
\be
\frac{\langle q| (p_3-x)\circ 3|q\rangle}{ \langle 43\rangle \langle 12\rangle \langle 23\rangle}
\left( \langle q|(p_1-x)\circ 1|2\rangle \langle q3\rangle - \langle q|(p_2-x)\circ (1+2)|q\rangle\langle 23\rangle \right).
\ee
We can replace $p_1$ by $p_4$ and $1$ by $(1+2)$ in the first term, and $p_2$ by $p_4$ in the second term. This gives for the expression in the brackets
\be
\langle q|(p_4-x)\circ (1+2)|2\rangle \langle q3\rangle - \langle q|(p_4-x)\circ (1+2)|q\rangle\langle 23\rangle = \langle q|(p_4-x)\circ (1+2)|3\rangle \langle q2\rangle.
\ee
This can be written as
\be\label{app-2}
\langle q|(p_4-x)\circ (1+2)|3\rangle \langle q2\rangle= \langle q|(p_4-x)\circ (1+2+3)|3\rangle \langle q2\rangle
= -\langle q|(p_4-x)|4] \langle 43\rangle \langle q2\rangle.
\ee
Given that we can replace here $p_4$ by $p_3$, it is clear that the terms ${\mathcal M}_{13}, {\mathcal T}^{12}_1, {\mathcal T}^{41}_3, {\mathcal B}^3$ cancel each other.

For the denominator $(p_2-x)^2(p_4-x)^2$ there are only two contributing terms ${\mathcal T}^{12}_2$ and ${\mathcal B}^{12}$, which directly cancel each other. For the denominator $(p_3-x)^2(p_4-x)^2$ we have the terms ${\mathcal M}_{23}, {\mathcal T}^{12}_3, {\mathcal T}^{23}_1, {\mathcal B}^4$ contributing, and the cancelation here is similar to the one encountered in the case $(p_2-x)^2(p_3-x)^2$.

Therefore the total integrand, as a sum of box, four triangles, two internal bubbles and eight external bubbles vanishes. Pictorially, this can be represented as 
\be
\begin{gathered}
\begin{fmfgraph*}(60,45)
\fmfbottom{i1,i2}
\fmftop{o1,o2}
\fmf{photon,tension=3}{i1,v1}
\fmf{photon,tension=3}{i2,v2}
\fmf{photon,tension=3}{o1,v3}
\fmf{photon,tension=3}{o2,v4}
\fmf{photon}{v3,v3'}
\fmf{vanilla}{v3',v1}
\fmf{photon}{v1,v1'}
\fmf{vanilla}{v1',v2}
\fmf{photon}{v2,v2'}
\fmf{vanilla}{v2',v4}
\fmf{photon}{v4,v4'}
\fmf{vanilla}{v4',v3}
\fmflabel{}{o1}
\fmflabel{}{o2}
\fmflabel{}{i1}
\fmflabel{}{i2}
\fmflabel{}{v1'}
\fmflabel{}{v3'}
\fmflabel{}{v2'}
\fmflabel{}{v4'}
\end{fmfgraph*}\nonumber
\end{gathered}~~+~~
4\times\begin{gathered}
\begin{fmfgraph*}(70,40)
\fmfbottom{i1,i2}
\fmftop{o1,o2}
\fmf{photon,tension=3}{i1,v1,o1}
\fmf{photon,tension=3}{i2,v3}
\fmf{photon,tension=3}{o2,v4}
\fmf{vanilla}{v1,v1'}
\fmf{photon}{v1',v2}
\fmf{photon}{v2,v2'}
\fmf{vanilla}{v2',v4}
\fmf{photon}{v4',v4}
\fmf{vanilla}{v4',v3}
\fmf{photon}{v3',v3}
\fmf{vanilla}{v3',v2}
\fmflabel{}{o1}
\fmflabel{}{o2}
\fmflabel{}{i1}
\fmflabel{}{i2}
\fmflabel{}{v1'}
\fmflabel{}{v2'}
\fmflabel{}{v3'}
\fmflabel{}{v4'}
\end{fmfgraph*}
\end{gathered}~~+~~
2\times \begin{gathered}
\begin{fmfgraph*}(70,30)
     \fmftop{i1,i3,i2}
     \fmfbottom{o1,o3,o2}
     \fmf{photon}{i1,v2,o1}
     \fmf{vanilla}{v2,v1'}
     \fmf{photon}{v1',v1}
     \fmf{photon}{v1,i3}
     \fmf{vanilla}{i3,v4}
     \fmf{photon}{v4,o3}
     \fmf{vanilla}{o3,v1}
     \fmf{photon}{v4,v3}
     \fmf{vanilla}{v3,v3'}
     \fmf{photon}{i2,v3',o2}
\end{fmfgraph*}
\end{gathered}~~+~~
4\times \begin{gathered} 
\begin{fmfgraph*}(70,30)
     \fmfleft{i}
     \fmfright{o}
     \fmfbottom{v}
     \fmftop{u}
     \fmf{photon}{i,v1}
     \fmf{photon}{v1,u}
     \fmf{vanilla}{u,v2}
     \fmf{photon}{v2,v}
     \fmf{vanilla}{v,v1}
     \fmf{photon}{v2,v3}
     \fmf{vanilla}{v3,o}
     \fmfblob{.15w}{o}
\end{fmfgraph*}
\end{gathered}~~=0
\ee

\end{fmffile}

\end{document}